\newcommand{\ud}{\rm d}
\newcommand{\un}{~\mathrm}

\newcommand{\eg}{{\em e.g. }}

\documentclass[nofootinbib,showpacs,preprintnumbers,amsmath,amssymb]{revtex4-1}

\usepackage[english,francais]{babel}
\usepackage[T1]{fontenc} 
\usepackage{graphicx}
\usepackage{wasysym}
\usepackage{multirow}
\usepackage{makecell}
\usepackage{url}

\begin{document}


\title{Crack growth in heterogeneous brittle solids: Intermittency, crackling and induced seismicity}

\author{Jonathan Bar{\'e}s$^{1}$ and Daniel Bonamy$^{2}$}

\address{$^{1}$Laboratoire de M{\'e}canique et G{\'e}nie Civil, Universit{\'e} de Montpellier, CNRS, 163 rue Auguste Broussonnet, 34090 Montpellier, France\\
$^{2}$Service de Physique de l'Etat Condens{\'e}, CEA, CNRS, Universit{\'e} Paris-Saclay, CEA Saclay 91191 Gif-sur-Yvette Cedex, France}




\begin{abstract}
Crack growth is the basic mechanism leading to the failure of brittle materials. Engineering addresses this problem within the framework of continuum mechanics, which links deterministically the crack motion to the applied loading. Such an idealization, however, fails in several situations and in particular cannot capture the highly erratic (earthquake-like) dynamics sometimes observed in slowly fracturing heterogeneous solids. Here, we examine this problem by means of innovative experiments of crack growth in artificial rocks of controlled microstructure. The dynamical events are analyzed at both global and local scales, from the time fluctuation of the spatially-averaged crack speed and the induced acoustic emission, respectively. Their statistics are characterized and compared with the predictions of a recent approach mapping fracture onset with the depinning of an elastic interface. Finally, the overall time-size organization of the events are characterized to shed light on the mechanisms underlying the scaling laws observed in seismology.
\end{abstract}

\maketitle

\section{Introduction and background}

Damage and failure are central to many fields, from civil to aerospace engineering, from nano- to Earth-scales. Yet, they remain difficult to anticipate: Stress enhancement at defects makes the behavior observed at the macroscopic scale extremely dependent on the presence of material inhomogeneities down to very small scales. As a consequence, in heterogeneous brittle solids upon slowly varying external loading, the failure processes are sometimes observed to be erratic, with random cascades of microfracturing events spanning a variety of scales. Such dynamics are \eg revealed by the acoustic noise emitted during the failure of various solids \cite{petri1994_prl,garcimartin97_prl,davidsen05_prl,baro13_prl} and, at much larger scale, by the seismic activity going along with earthquakes \cite{bak02_prl,corral04_prl}. Generic features in the field are the existence of scale-free statistics for individual microfracturing/acoustic/seismic events (see \cite{bonamy2009_jpd} for a review) and the non-trivial organization of the event sequences into characteristic aftershock sequences obeying specific laws initially derived in seismology (see \cite{bonamy2009_jpd} for a review).  

For brittle solids under tension, the difficulty is tackled by reducing the problem down to that of the destabilization and subsequent growth of a single pre-existing crack \cite{Bonamy17_crp}. Linear Elastic Fracture Mechanics (LEFM) provides a powerful framework to address this so called situation of nominally brittle fracture, and links deterministically crack dynamics to applied loading\cite{lawn93_book}. Still, such a continuum approach fails in some situations. In particular, the crack growth is sometimes observed \cite{maloy06_prl,Marchenko06_apl,Astrom06_pl,Kovoisto07_prl,Stojanova14_prl} to be erratic, made of random and local front jumps -- avalanches -- whose statistics share some of the scale-free features mentioned above. This so-called crackling dynamics can be interpreted by mapping the in-plane motion of the crack front to the problem of a long-range (LR) elastic interface propagating within a two-dimensional random potential \cite{Schmittbuhl95_prl,ramanathan97_prl}, so that the driving force self-adjusts around the depinning threshold \cite{bonamy2008_prl}. This approach reproduces quantitatively many of the statistical features observed in the simplified 2D experimental configuration of an interfacial crack driven along a weak heterogeneous plate \cite{bonamy2008_prl,Laurson13_natcom,Ponson17_pre}. Still, whether or not this approach allows describing the bulk fracture of real three-dimensional solids remains an open question (see \cite{bares14_prl} for preliminary work in this context). Beyond their individual scale-free features, whether or not the events get organized into the characteristic aftershock sequences of seismology in this more tractable single crack problem is an important question (see \cite{grob09_pag,bares2018_natcom} for preliminary works).    

The work gathered here aims at filling this gap. We designed a fracture experiment which consists in driving a tensile crack throughout an artificial rock of tunable microstructure. At slow enough driving speed, the crack dynamics displays an irregular burst-like dynamics. The fluctuations of instantaneous crack speed and mechanical energy release are both monitored and used to characterize the crackling dynamics at the continuum-level (global) scale. The induced acoustic events are recorded and provide information at the local scale. The so-obtained experimental data are contrasted with the crackling features predicted by the depinning approach at both global and local scales. Beyond their individual statistics the time-energy organization is analyzed in a similar way to that developed in statistical seismology.  

\section{Material \& Methods}

\begin{figure}[!h]
\centering\includegraphics[width=\columnwidth]{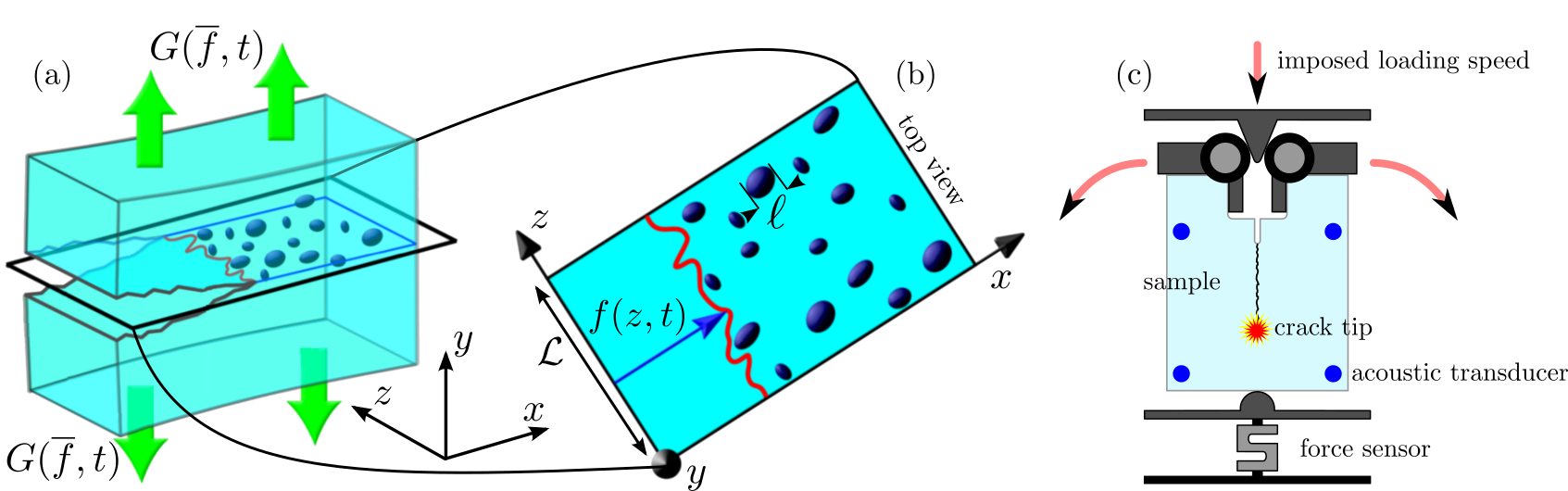}
\caption{(a) This sketch depicts a nominally brittle crack propagating in an heterogeneous solid in opening mode due to a prying forcing quantified by the elastic energy release rate $G(\overline{f},t)$. (b) The time evolution of the crack front (red solid line) projected onto the mean crack plane $(x,z)$ is described by the function $f(z,t)$. The sample width is $\mathcal{L}$ and the characteristic heterogeneity size is $\ell$. (c) Sketch of the experimental fracture set-up. A model rock made of sintered polymer bead is fractured by means of a wedge-splitting geometry, by pushing at constant speed a triangular wedge into a rectangular notch cut out on the sample. This allows driving a slow stable crack in tension (red arrows). During the crack growth, the propagation is monitored by eight acoustic transducers (four in the front four in the back) and a global force sensor.}
\label{fig_method}
\end{figure}

\subsection{Theoretical \& numerical aspects}

The continuum framework of LEFM addresses the problem of a straight slit crack embedded in an homogeneous solid. Crack motion is governed by the balance between the amount of mechanical energy released by the solid as the crack propagates over a unit length, $G$, and the fracture energy, $\Gamma$, which is the energy dissipated in the fracture process zone to create two new fracture surfaces of unit area \cite{lawn93_book}. In the standard LEFM framework, $G$ depends on the imposed loading and specimen geometry and $\Gamma$ is a material constant. For a slow enough motion,  crack speed $v$ is given by:

\begin{equation}
	\dfrac{1}{\mu} v=G - \Gamma,	
\end{equation}

\noindent where $\mu$ is the crack front mobility. In a perfect linear elastic material (and in the absence of any environmental effect such as stress corrosion for instance), $\mu$ can be related to the Rayleigh wave speed $c_R$ via $\mu= c_R / \Gamma$. For a viscoelastic material like the polystyrene used here, viscoelasticity effects are not negligible and $\mu$ is expected to be much smaller.

The depinning approach explicitly introduces the microstructure disorder (see Fig. \ref{fig_method}) by adding a stochastic term in the local fracture energy: $\Gamma(x,y,z)=\overline{\Gamma}+\eta(x,y,z)$. Here and thereafter,  $x$, $y$ and $z$ axis are respectively oriented along the growth direction, tensile loading direction, and front direction, as shown in Fig. \ref{fig_method}. This induces in-plane and out-of-plane distortions of the front which, in turn, generates local variations in $G$. 
As such, the problem is {\em a-priori} 3D; however, to first order, it can be decomposed into two independent effective 2D problems: an equation of motion with describes the dynamics of the in-plane projection of the crack line and an equation of trajectory which describes the $x$ evolution of the out-of-plane roughness -- $x$ being the analog of time (see \cite{bares2014_ftp} for details). The underlying reasons are: (i) The out-of-plane corrugations are logarithmically rough \cite{larralde1995_epl,ramanathan97_prl,bares2014_ftp} and $\vec{v}$ and $\eta(x,y,z)$ reduces to their in-plane projections at large scales; (ii) to first order, to the first order, the variations of $G$ depend on the in-plane front distortion only \cite{movchan1998_ijss}. One can then use Rice's analysis \cite{rice1985_jam,gao1989_jam} to relate the local value $G(z,t)$ of energy release to the in-plane projection of the front shape, $f(z,t)$ (Fig. \ref{fig_method}(b)):

\begin{align}
	& G(z,t) =\overline{G}(1+J(z,\lbrace f \rbrace)), \label{eq_mod0}\\
    & \mathrm{with} \, J(z,\lbrace f \rbrace)=\dfrac{1}{\pi} \times pp \int_{\text{front}}{\dfrac{f(\zeta,t)-f(z,t)}{(\zeta-z)^2}d\zeta}, \nonumber  	
\end{align}

\noindent where $pp$ denotes the \textit{principal part} of the integral. Note that the long-range kernel $J$ is more conveniently defined by its $z$-Fourier transform $\hat{J}(q)=-\vert q \vert \hat{f}$. $\overline{G}$ denotes the energy release rate that would have been used in the standard LEFM description, after having coarse-grained the microstructure disorder and replaced the distorted front by a straight one at the effective position $\overline{f}(t)=\langle f(z,t) \rangle_z$ obtained after having averaged over the specimen thickness. Once injected in the equation of motion, this yields:

\begin{equation}
	\dfrac{1}{\mu} \dfrac{\partial f}{\partial t}=F(\overline{f},t)+\overline{\Gamma}J(z,\lbrace f \rbrace)+\eta(z,x=f(z,t)), 	
	\label{eq_mod1}
\end{equation}

\noindent where $F(\overline{f},t)=G(\overline{f},t)-\overline{\Gamma}$ is the loading. The random term $\eta(z,x)$ is characterized by two main quantities, the noise amplitude defined as $\tilde{\Gamma}=\langle \eta^2(z,x)\rangle^{1/2}_{x,z}$ and the spatial correlation length $\ell$ over which the correlation function $C(x,z)=\langle \eta(x_0+x,z_0+z)\eta(x_0,z_0)\rangle_{x_0,z_0}$ decreases.

We consider now situations of stable growth -- both in terms of dynamics and trajectory. These are encountered in systems of geometry making $G$ decrease with crack length, keeping the T-stress negative and loaded externally by imposing time-increasing displacements \cite{bonamy2008_prl,bares2014_ftp}. Then, $F(\overline{f},t)$ writes \cite{bares13_prl}:

\begin{equation}
	F(\overline{f},t)=\dot{G}t-G'\overline{f}
	\label{eq_mod2}
\end{equation}

\noindent where $\dot{G}=\partial G/\partial t$ (driving rate) and $G'=-\partial G/\partial\overline{f}$ (unloading factor) are positive constants. Equations \ref{eq_mod1} and \ref{eq_mod2} provide the equation of motion of the crack line. It is convenient to introduce dimensionless time, $t\rightarrow t/(\ell/\mu\overline{\Gamma})$, and space, $\{x,z,f\}\rightarrow \{x/\ell,z/\ell,f/\ell\}$ to reduce the number of parameters from seven to four:

\begin{equation}
	\dfrac{\partial f}{\partial t}=ct - k \overline{f}+\overline{\Gamma}J(z,\lbrace f \rbrace)+\eta(z,f(z,t)), 
\label{eqLine}	
\end{equation} 

\noindent where $c=\dot{G}\ell/\mu\overline{\Gamma}^2$ is the dimensionless loading speed, $k=G'\ell/\overline{\Gamma}$ is the dimensionless unloading factor. The two other parameters are the system size $N$ (in $\ell$ unit) and the dimensionless noise amplitude $\tilde{\Gamma} \rightarrow \tilde{\Gamma}/ \overline{\Gamma}$.

In the following, all these parameters were fixed to values ensuring a clear crackling dynamics \cite{bares13_prl}, with scale-free statistics ranging over a wide number of decades: $c=2 \times 10^{-6}$,  $k=10^{-4}$, $\tilde{\Gamma}=1$ and $N=1024$. The front line is discretized along $z$, $f(z,t)=f_z(t)$ with $z \in \{1,..., N\}$. The time evolution of $f_z(t)$ is obtained by solving Eq. \ref{eqLine} via a fourth order Runge-Kutta scheme (discretization time step: $dt=0.1$), as in \cite{bares13_prl,bares2014_ftp}. The space-time dynamics $f(z,t)$ is obtained. Its time derivative gives the local front speed, $v(z,t)=df_z(t)/ dt$ and the spatially-averaged crack speed is deduced (Figure \ref{fig_display}(a)): 

\begin{equation}
\overline{v}(t)=\frac{1}{N}\sum_{z=1}^{N} v(z,t)
\end{equation} 

\noindent As we will see in section \ref{Sec:definition}, the global events will be identified with the bursts in $\overline{v}(t)$, while the local ones will be dug out from the space-time maps $v(z,t)$. The movie provided as electronic supplementary material shows the evolution of both $v(z,t)$ and $\overline{v}(t)$.

\subsection{Experimental aspects}

The fracture experiments presented here were carried out on an home-made artificial rock obtained by sintering polystyrene beads. 
The sintering procedure is detailed in \cite{cambonie2015_pre,bares2018_natcom} and summarized herein. 
First, a mold filled with monodisperse polystyrene beads (Dynoseeds from Microbeads SA, diameter $d$) is heated up to $90$\% of the temperature at glass transition, $T=105^\circ$C. 
Then, the  mold is gradually compressed up to a prescribed pressure, $P$, by means of an electromechanical loading machine, while keeping $T=105^\circ$C. 
Both $P$ and $T$ are then kept constant for one hour to achieve the sintering. Then, the system is unloaded and cooled down to ambient temperature at a rate slow enough to avoid residual stress ($\sim 8$ hours to cool down from $T$ to room temperature). This procedure provides a so-called artificial rock of homogeneous microstructure, the porosity and length-scale of which are set by the prescribed values $P$ and $d$ \cite{cambonie2015_pre}. Note that the formation of natural rocks are much more complex and cannot be approached by a process as the one used here. However, our model materials share two important features of the simplest rocks (sandstone for instance): They are composed of small cemented grains and the cracks propagate in a brittle manner between these grains. In the experiments reported here, $d=583~\mu\mathrm{m}$ and $P$ is large enough (larger than $1\un{MPa}$) so that a dense rock is obtained, with no porosity. 
It breaks in a nominally brittle manner, by the propagation of a single inter-granular crack in between the sintered grains. 
The disordered nature of the grain joint network yields small out-of-plane deviations -- roughness --, the statistics of which has been analyzed in \cite{cambonie2015_pre}. 
These out-of-plane deviations, in turn, result in small variations in the landscape of effective toughness (term $\eta(x,z)$ in Eq. \ref{eqLine}). 
The typical length-scale to be associated with this quenched disordered toughness, hence, is set by $d$ \cite{bares2018_natcom,cambonie2015_pre}.

In the so-obtained materials, stable cracks were driven by means of the wedge splitting fracture test depicted in Fig. \ref{fig_method}(c). Parallelepipedic samples of length $140\un{mm}$ (along $x$), width $125\un{mm}$ (along $y$), and thickness $15\un{mm}$ (along $z$) are first machined. A rectangular notch is then cut out on one of the two lateral $(y-z)$ edges and an initial seed crack ($10$~mm long) is introduced in the middle of the cut with a razor blade. A triangular wedge (semi-angle $15^{\circ}$) is then pushed into this rectangular notch at a constant speed $V_{\text{wedge}}=16$~nm/s (Fig. \ref{fig_method}(c)). When the applied loading is large enough, the seed crack destabilizes and starts growing. During the experiment, the force $F(t)$ applied by the wedge is monitored via a S-type Vishay cell force (acquisition rate of $50$~kHz, accuracy of $1$~N), and the instantaneous specimen stiffness $\kappa(t) = F(t)/V_{\text{wedge}} \times t$ is deduced. Such a wedge splitting arrangement also ensure stable crack paths: The compression along x induced by the wedge (vertical axis on Fig. 1c) produces a negative T-stress \cite{seitl2011_cs} and, hence, encourages  the crack to stay in the vicinity of the symmetry plane of the specimen ($y=0$) at large scales \cite{cotterell1980_ijf}.

Two go-between steel blocks placed between the wedge and the specimen limit parasitic mechanical dissipation and ensure the damage and failure processes to be the dominating dissipation source for mechanical energy in the system (see \cite{bares14_prl,bares2018_natcom} for more details). As a result, both the instantaneous elastic energy stored in the specimen, $\mathcal{E}(t)$, and instantaneous crack length (spatially averaged over specimen thickness), $\overline{f}(t)$, can be determined with very high resolution (see \cite{bares14_prl} for details). Indeed, in a linear elastic material, $\mathcal{E}(t)=\frac{1}{2}F^2(t)/\kappa(t)$ and, for a prescribed geometry, $\kappa$ is a function of $\overline{f}$ only. The reference curve $\kappa$ \textit{vs.} $\overline{f}$ curve was then computed in our geometry by finite element calculations (Cast3M software), and used to infer the spatially-averaged crack position at each time step: $\overline{f}(t)=\kappa^{-1}(F(t)/V_{\text{wedge}} \times t)$. Time derivation of $\overline{f}(t)$ and $-\mathcal{E}(t)$ provides the instantaneous crack speed, $\overline{v}(t)$ and mechanical power released, $\mathcal{P}(t)$ (Fig. \ref{fig_display}(f)). Both quantities were found to be proportional \cite{bares14_prl}. This actually results from the nominally brittle character of the specimen fracture, so that the mechanical energy release rate per unit length, $G=-\ud \mathcal{E}/\ud \overline{f}=\mathcal{P}(t)/\overline{f}(t)$, is equal at each time step to the fracture energy, $\Gamma$, which is a material constant. For the artificial rocks considered here: $\Gamma =100\un{J/m}^2$ \cite{bares14_prl}.   

Note finally that, in addition to $\overline{f}(t)$ and $\mathcal{P}(t)$, the acoustic emission was collected at eight different locations via eight broadband piezoacoustic transducers (see \cite{bares2018_natcom} for details). The signals were preamplified, band-filtered, and recorded via a PCI-2 acquisition system (Europhysical Acoustics) at $40$~MSamples/s. An acoustic event (AE), $i$, is defined to start at the time $t^{start}_i$ when the preamplified signal $\mathcal{V}(t)$ goes above a prescribed threshold ($40$~dB), and to stop when $\mathcal{V}(t)$ decreases below this threshold at time $t^{end}_i$. The minimal time interval between two successive events is $402$~$\mu$s. This interval breaks down into two parts: The hit definition time (HDT) of $400\,\mu$s and the the hit lockout time (HLT) of $2\,\mu$s. The former sets the minimal interval during which the signal should not exceed the threshold after the event initiation to end it and the latter is the interval during which the system remains deaf after the HDT to avoid multiple detections of the same event due to reflexions.

The wave speed in our model rocks was measured to be  $c_W=2048\un{m/s}$ and the emerging waveform frequency, $\nu$, ranges from $40$ to $130\un{kHz}$ depending on the considered event. This yields typical wavelengths $\lambda=c_W/\nu = 1.5-5\un{mm}$. Such wavelengths are of the order of the specimen thickness and, as such, are conjectured to coincide with the resonant modes of the plate. We hence propose the following scenario: As a depinning event occurs and the front line jumps over an increment, an acoustic event is produced. The frequencies of the so-emitted pulse spans {\em a priori} from $\sim 40\un{kHz}$ (resonant modes) to few MHz (selected by the characteristic jump size, of the order of $d$). Due to the absorption properties of the material (a polymer, that is a viscoelastic material), the high frequency portion of the signal attenuates rapidly and only the lowest frequency part survives when the pulse reaches the transducers. 
   
Each so-detected AE is characterized by three quantities: occurrence time, energy and spatial location. The occurrence time is identified with the starting time $t^{start}_i$.  Its energy is defined as the squared maximum value of $V(t)$ between $t^{start}_i$ and $t^{end}_i$. In the scenario depicted above, indeed, the pulse duration is not correlated to the underlying depining event and the initial value is more relevant than the integral over the whole duration; we checked however that the results do not change if the event energy is defined as this integral \cite{bares2018_natcom}. The spatial location is obtained from the arrival time at each of the eight transducers. The spatial accuracy, here, is set by the typical pulse width $\lambda \simeq 5\un{mm}$.  

The movie provided as electronic supplementary material shows the synchronized evolution of both the continuum-level scale quantities and acoustic events as the crack is driven in our artificial rock. As in the numerical simulation the global events will be identified with the bursts in the signal $\overline{v}(t)$ (see next section). A priori, acoustic events are more connected to the local avalanches, but, as will be seen later in this manuscript, there is no direct mapping between the two. 

\section{On the different types of avalanches and their production rate}\label{Sec:definition}

The dynamics emerging in the above experiments and simulations are analyzed both at the global and local scale.  

\begin{figure}[!h]
\centering\includegraphics[width=1.05\columnwidth]{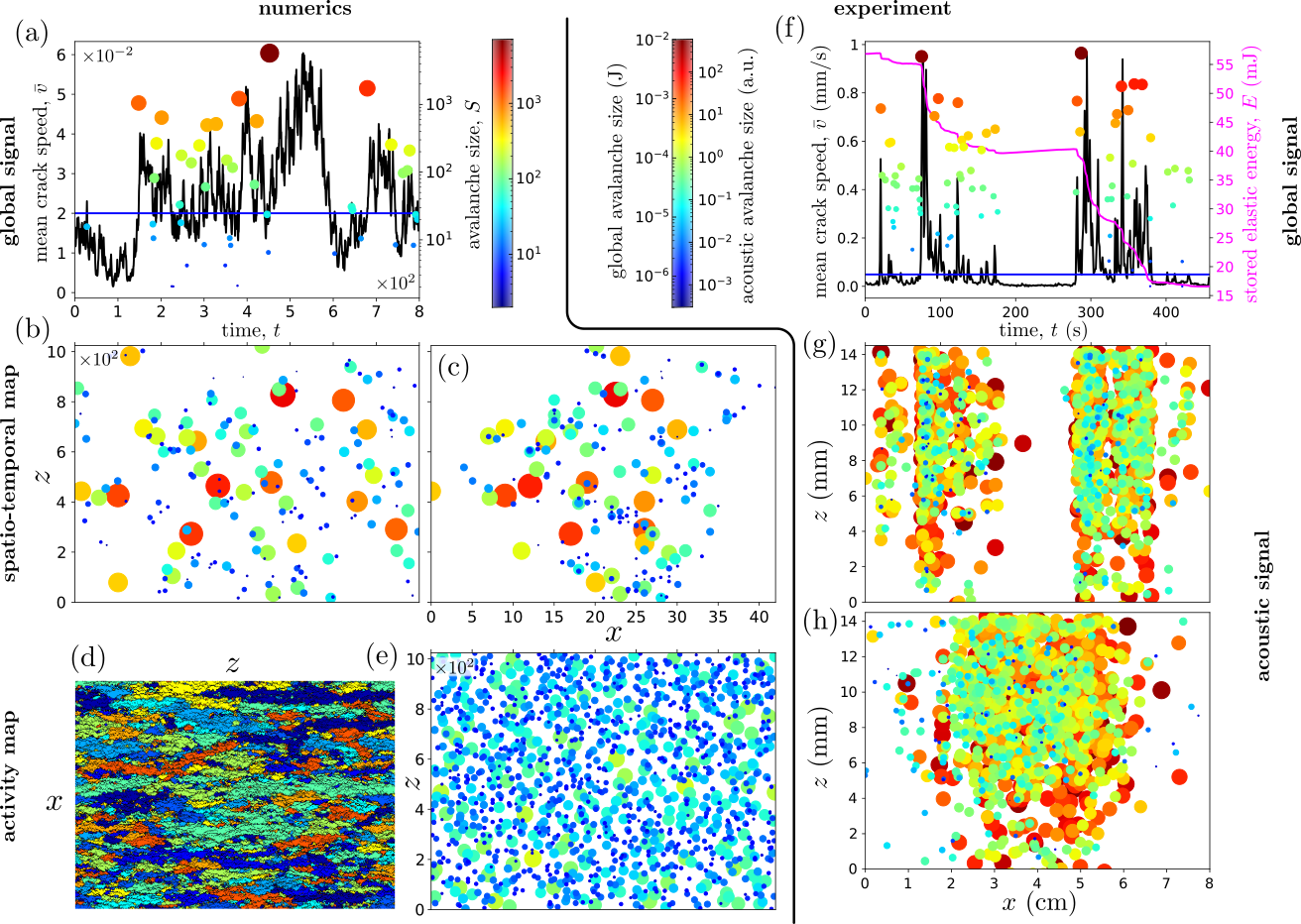}
\caption{(a): Evolution of the mean crack speed $\bar{v}$ in the numerical simulations. The blue horizontal line shows the avalanche detection threshold $v_{th}$ and the colored discs display the size of the detected avalanches according to the colorbar provided on the right. (b),(c): Position of the avalanches detected at the local scale on the $v(z,t)$ signal, in the spatio-temporal and spatial maps respectively. The disc color indicates the avalanche size. (d),(e): Avalanches detected on the activity map. Different colors stand for different avalanches in (d) while in (e) the color code indicates the avalanche size. (f): Evolution of the mean crack speed $\bar{v}$ in the experiment. The blue horizontal line shows the avalanche detection threshold $v_{th}$ and the colored discs display the size of the detected avalanches. The magenta curve shows the evolution of the elastic energy $E$ stored in the system. (g),(h): Position of the avalanches detected via the acoustic transducers, in the spatio-temporal and spatial maps respectively. The disc color indicates the avalanche size. The figures (a)-(e) on the left were obtained from the numerical simulations while the figures (f)-(h) on the right were obtained from experiments. In all figures,the disc radius is proportional to the logarithm of the avalanche size.}
\label{fig_display}
\end{figure}

Figures \ref{fig_display}(a) and \ref{fig_display}(f) display the time evolution of $\overline{v}$ for the simulation and experiment respectively. Erratic dynamics are observed, with sharp bursts corresponding to the sudden jumps of the crack front. These jumps are thereafter referred to as global avalanches or events. To dig them out, we adopt the standard procedure used for crackling signals \cite{sethna01_nature}; a threshold $v_{th}$ is prescribed and the avalanches are identified with the parts of the signal where $\overline{v}(t) \geq v_{th}$ (Fig. \ref{fig_display}(a) and (f)). The avalanche $i$ starts at time $t^{start}_i$ when the signal $\overline{v}(t)$ first rises above $v_{th}$, and subsequently ends at time $t_i^{end}$ when $\overline{v}(t)$ returns below this value. The position $x_i$ of this avalanche is defined as $x_i=\overline{f}(t^{start}_i)$. The avalanche size $S_i$, in the numerical case, is defined as the area swept by the crack front during the burst: $S_i=N \int_{t_i^{start}}^{t_i^{end}} (\overline{v}(t) - v_{th}) dt$. In the experimental case, $S_i$ is defined as the energy released during avalanche $i$: $S_i=\mathcal{E}(t_i^{end})-\mathcal{E}(t_i^{start})$. Let us recall here that this energy released is proportional to the area swept by the crack front during the event, and the proportionality constant is $\Gamma$ \cite{bares14_prl}. Examples of avalanches detected with this method are displayed in Fig. \ref{fig_display}(a) and (f).

In the numerical simulations, the jumps of the crack line can also be analyzed at the local scale, from the space-time evolution of $v(z,t)$. Two distinct methods are used to identify the avalanches. In both cases, special attention has been paid to take properly into account the periodic boundary conditions in the clustering methods. 

The first method, pioneered by \cite{tanguy1998_pre}, is a generalization of the procedure used to dig out the global avalanches. We consider the spatio-temporal map $v(z,t)$ and apply the same threshold $v_{th}$ as the one considered for global avalanches. The avalanches are then defined as the connected clusters, in the $(z,t)$ space, where $v(z,t)>v_{th}$. Avalanche $i$ starts at time $t^{start}_i$ defined as the first time where $v(z,t)>v_{th}$ in the considered cluster. It ends at $t^{end}_i$ which is the last time so that $v(z,t)>v_{th}$ in the same area. Avalanche size $S_i$ is given by the local area swept by $f(z,t)$ between $t^{start}_i$ and $t^{end}_i$. The 2D avalanche position; $(x_i,z_i)$; is defined such as $x_i=f(z_i,t^{start}_i)$ where $z_i$ is the first location (in $z$) where $f$ enters into the considered cluster at $t^{start}_i$ (see \cite{bares13_phd} for details). An example of the location of these local avalanches is shown in Fig. \ref{fig_display}(b) and (c).

The second method used here to identify the local avalanches was initially proposed by \cite{maloy06_prl}. It consists in building a space-space activity map,$W(x,z)$, from the time spent by the crack line at each location $(x,z)$. The inverse of this map provides a space-space cartography of local speeds, $V(x,z)=1/W(x,z)$. A threshold value, $V_{th}$, is then defined and the avalanches are identified with the clusters of connected points where $V(x,z) \geq V_{th}$. Such an activity map is shown in Fig. \ref{fig_display}(d). The avalanche size $S_i$ is given by the cluster area, its position $(x_i,z_i)$ is defined by that of its center of mass and its duration $D_i$ is the sum of the waiting times $W(x,z)$ over the considered cluster (\textit{cf}. \cite{bares13_phd} for details). Note that an accurate occurrence time cannot be attributed to the avalanche identified within this method.

The procedure described above to dig out avalanches at the local scales from the space-time dynamics of $v(z,t)$, unfortunately, cannot be applied to our experiments. Conversely, these local avalanches may be at the origin of the acoustic events recorded during our experiments. As such, these latter have been analyzed accordingly (Fig. \ref{fig_display}(h)).

The different methods presented above allow obtaining catalogs, for both local and global avalanches in the numerical and experimental experiments, which gathers different quantities: First the avalanche size $S_i$ and position $x_i$ along the crack propagation direction for all types of events. Considering local avalanches, their position $z_i$ along the crack is also measured. For all methods but the one based on activity map, starting and ending time, $t^{start}_i$ and $t^{end}_i$ are also determined; occurrence time, $t_i$ is then identified with $t^{start}_i$. The duration $D_i$ of each avalanche is deduced: $D_i=t^{end}_i-t^{start}_i$. The waiting time $\Delta t_i$ between two consecutive avalanches is computed as $\Delta t_i=t^{start}_{i+1}-t^{start}_i$. When the spatial location of the avalanche is obtained just like in the case of the local avalanches measured from the $v(z,t)$ map, we also define the jump $\Delta r_i$ between two consecutive avalanches as $\Delta r_i=\sqrt{(x_{i+1}-x_{i})^2+(z_{i+1}-z_{i})^2}$. Table \ref{tab_method} synthesizes the five types of avalanches considered here (two for the experiment, three for the simulation) and the quantities collected in their respective catalogs.

\begin{table}[!h]
\begin{center}
\begin{tabular}{|c|c|c|c|c|c|c|c|}
\hline 
  & $S$ & $t$ & $D$ & $x$ & $z$ & $\Delta t$ & $\Delta r$ \\ 
\Xhline{3\arrayrulewidth}
numerical $\overline{v}(t)$ signal & $\times$ & $\times$ & $\times$ & $\times$ &   & $\times$ &   \\ 
\hline 
experimental $\overline{v}(t)$ signal & $\times$ & $\times$ & $\times$ & $\times$ &   & $\times$ &   \\
\Xhline{3\arrayrulewidth}
numerical spatio-temporal $v(z,t)$ map & $\times$ & $\times$ & $\times$ & $\times$ & $\times$ & $\times$ & $\times$ \\ 
\hline 
numerical activity map $W(x,z)$ & $\times$ &   & $\times$ & $\times$ & $\times$ &   &   \\ 
\hline 
experimental acoustic signal & $\times$ & $\times$ &   & $\sim$ & $\sim$ & $\times$ & $\sim$  \\ 
\hline 
\end{tabular}
\caption{Synthesis of the different types of avalanches defined here and associated catalogs. The first two columns are to be associated with the global avalanches while the three formers are connected to the local avalanches. $S$, $t$, $D$ $x$, $z$, $\Delta t$, $\Delta r$ denote size, occurrence time, duration, position along growth direction, position along crack front, inter-event time and inter-event distance, respectively. $\times$ denotes accurate measurements while $\sim$ denotes coarse ones.}
\label{tab_method} 
\end{center}
\end{table}

\begin{figure}[!h]
\centering\includegraphics[width=0.8\columnwidth]{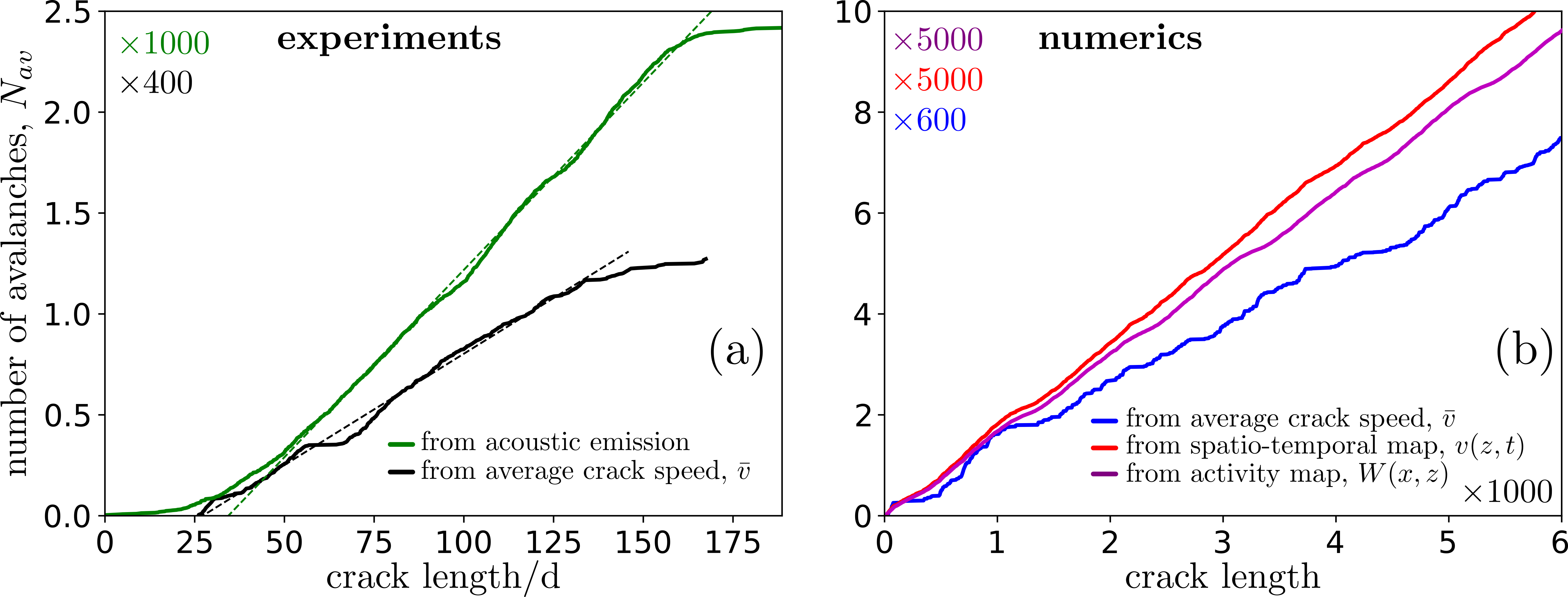}
\caption{Cumulative number of events as a function of crack length for experiments (left) and simulations (right). 
The different curves stand for the different types of avalanches: acoustic events (green, panel a), events detected on the experimental $\overline{v}(t)$ signal (black, panel a), events detected on the numerical $\overline{v}(t)$ signal (blue, panel b), events detected on the numerical spatio-temporal map $v(z,t)$ (red, panel b), events detected on the activity map $W(x,z)$ (purple, panel b). All curves have been fitted linearly (dashed lines), the obtained density of events $s$ are: $s_{ea}=18.76\pm0.02$~avl./$d$, $s_{eg}=4.51\pm0.02$~avl./$d$, $s_{na}=8.01$~avl/s.u. (space unit), $s_{nl}=8.67$~avl/s.u. and $s_{ng}=0.71$~avl/s.u..}
\label{fig_space_density}
\end{figure}

Figure \ref{fig_space_density} displays the cumulative number of avalanches as a function of the length traveled by the crack, for all types of events. In all cases, the number of events linearly increases with crack length. For acoustic avalanches (Fig. \ref{fig_space_density}(a)), this has been interpreted by stating that the production rate of acoustic events is simply given by the number of heterogeneities met by the crack front as it propagates over a unit length \cite{bares2018_natcom}; this suggests a density of events $s_{ea}\sim \mathcal{L}/d^2$, which is of the order of the measured value\footnote{Here and thereafter, subscript $ea$ stands for 'experiment acoustic'.} ($s_{ea}=18.76$~avl/$d$). Still the different ways to define avalanches for the same sample induce rates that are orders of magnitude different from each other: it goes from $4.51$~avl/$d$ for the global speed signal, to $18.76$~avl/$d$ for the acoustic signal, in the experiment (Fig. \ref{fig_space_density}(a)); and from $0.71$~avl/s.u. (space unit) for the global speed signal to $8.67$~avl/s.u. for avalanche detected on the spatio-temporal map, in the simulation (Fig. \ref{fig_space_density}(b)). This suggests that avalanches detected on different local or global signals are not easy to map with each others. However the very close avalanche rates for both local detection methods on $v(z,t)$ and $W(x,z)$ ($s_{na}=8.01$~avl/s.u.) suggests that avalanches are similar\footnote{Here and thereafter, subscript $na$ stands for 'numerics activity'.}.

\section{Statistical features of individual events}

\begin{figure}[!h]
\centering\includegraphics[width=\columnwidth]{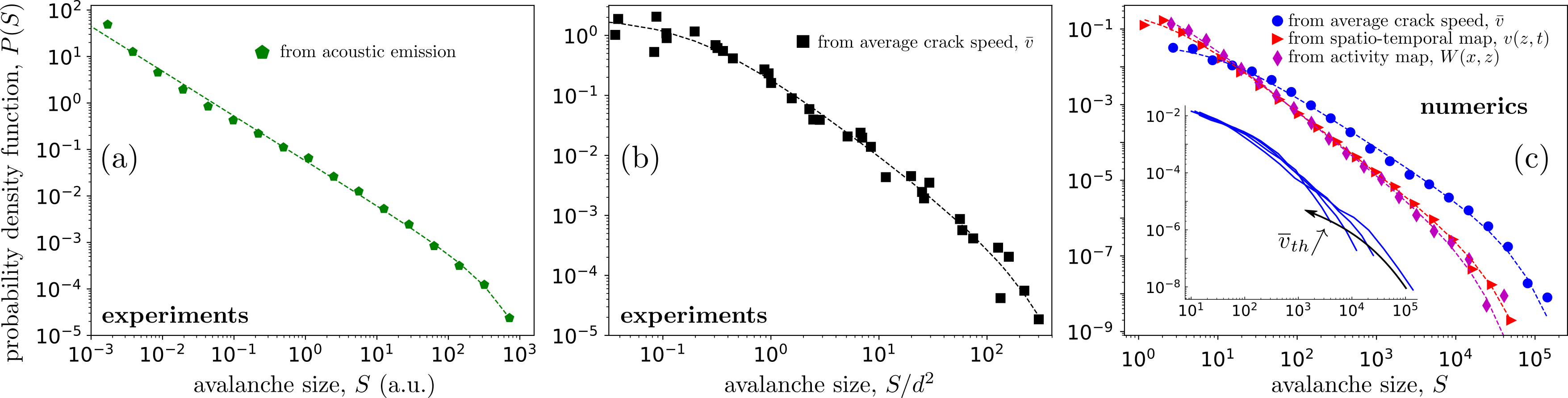}
\caption{Distribution of individual event size $P(S)$ for experiments (panel a and b) and simulations (panel c). The different curves stand for the different types of avalanches: acoustic events (green $\pentagon$, panel a), global events detected on the experimental $\overline{v}(t)$ signal (black $\Square$, panel b), global events detected on the numerical $\overline{v}(t)$ signal (blue $\Circle$, panel c), local events detected on the numerical spatio-temporal map $v(z,t)$ (red $\rhd$, panel c), and local events detected on the activity map $W(x,z)$ (purple $\Diamond$, panel c). All curves have been fitted using Eq. \ref{eq_PS} (dashed lines). The obtained fitting parameters are: $S_{min}^{ea}<<10^{-3}$, $S_{max}^{ea}=4.93\times10^{2}\pm0.11\times10^{2}$ and $\beta_{ea}=0.96\pm0.03$; $S_{min}^{eg}=0.20\pm 0.09$, $S_{max}^{eg}=1,90\times10^{2}\pm0.72\times10^{2}$ and $\beta_{eg}=1.35\pm0.1$; $S_{min}^{ng}=9.5\pm 4.4$, $S_{max}^{ng}=3.9\times10^{4}\pm0.7\times10^{4}$ and $\beta_{ng}=1.30\pm0.03$; $S_{min}^{nl}=2.04\pm0.5$, $S_{max}^{nl}=1.8\times10^{4}\pm0.2\times10^{4}$ and $\beta_{nl}=1.62\pm0.03$; $S_{min}^{na}=1.17\pm0.80$, $S_{max}^{na}=1.10\times10^{4}\pm0.18\times10^{4}$ and $\beta_{na}=1.66\pm0.05$. The inset in panel c shows $P(S)$ obtained from the numerical $\overline{v}(t)$ signal, for different $\overline{v}_{th}$. This parameter only have an effect on the upper cut-off. In panel b, the points are obtained by superimposing data from different avalanche detection threshold $\overline{v}_{th}$. The size is then scaled by the bead size $d$.}
\label{fig_richter}
\end{figure}

We first look at the statistics of individual events. In this context a generic feature common to crackling systems is the observation of scale-free statistics and scaling laws, characterized by well defined exponents \cite{sethna01_nature}. We first compare the statistics of avalanche size $S$ as obtained for the different definitions of avalanches. As presented in Fig. \ref{fig_richter}, in all cases and both in experiments and numerics, the statistics is scale-free; the probability density function (PDF) $P(S)$ follows a power-law spanning over several decades. More particularly, $P(S)$ is well fitted by:

\begin{equation}
P(S) \sim \frac{e^{-S/S_{max}}}{(1+S/S_{min})^{\beta}} ,
\label{eq_PS}
\end{equation}

\noindent where $S_{min}$ and $S_{max}$ are the lower and upper cut-offs respectively and $\beta$ is the exponent of this gamma law. Equation \ref{eq_PS} is reminiscent of the Gutenberg-Richter law for earthquake energy\footnote{Note however that, contrary to what is presented in Fig. \ref{fig_richter}, the energy distribution observed in seismology often takes the form of a pure power-law. As such, the earthquake energy $E$ -- analog to the size here -- is more commonly quantified by its magnitude, which is linearly related to the logarithm of the energy \cite{kanamori1977_jgr}: $\log_{10}(E) = 1.5M + 11.8$. The energy distribution is then presented via the classical Gutenberg-Richter frequency-magnitude relation: $\log_{10}(N(M)) = a - bM$, where $N(M)$ is the number of earthquakes per year with magnitude larger than $M$ and $a$ and $b$ are constants. This having been defined, the $b$-value relates to the exponent $\beta$ involved in Eq. \ref{eq_PS} via: $\beta = b/1.5 + 1$.} \cite{gutenberg44_bssa,gutenberg56_bssa}

Figure \ref{fig_richter}(a) does not reveal any smooth lower cut-off $S_{min}$ on the acoustic event (at least larger than the value $10^{-4}$ corresponding to the sensitivity of the acquisition system). The acoustic exponent is $\beta_{ea}=0.96\pm0.03$ \cite{bares2018_natcom}. This exponent is significantly lower than the one to be associated with the size distribution of global avalanches, displayed in Fig. \ref{fig_richter}(b): $\beta_{eg}=1.35\pm 0.1$ \footnote{Here and thereafter, subscript $eg$ stands for 'experiment global'.}. This value was found to decrease as $\overline{v}$ increases \cite{bares14_prl}, but always remains significantly larger than $\beta_{ea}$. As emphasized in \cite{bares2018_natcom}, there is no one-to-one correspondence between acoustic and global events; in particular, the number of the former is much larger than that of the latter (see end of section \ref{Sec:definition} and Fig. \ref{fig_space_density}). 

Concerning global avalanches, the size distribution are similar in the experiments and simulations: Within the error-bars, the exponents are the same: $\beta_{eg}=1.35\pm0.1$ and $\beta_{ng}=1.30\pm0.03$ (Fig. \ref{fig_richter}(c))\footnote{Here and thereafter, subscript $ng$ stands for 'numerics global'.}. These exponents are also in agreement with the one predicted for the long range depinning transition $\beta_g= 1.28$ \cite{bonamy2009_jpd,ledoussal09_pre}. 

At local scale, the observed exponents are significantly higher. Avalanches dug out from the spatio-temporal map reveal an exponent\footnote{Here and thereafter, subscript $nl$ stands for 'numerics local'.} $\beta_{nl}=1.62\pm0.03$ while those identified in the activity map are characterized by $\beta_{na}=1.66\pm 0.05$ (Fig. \ref{fig_richter}(c)). The similarity between the two, again, suggests that these two procedures to identify avalanches at the local scale are equivalent. Note that these two exponents are compatible with the values observed in earlier simulations \cite{bonamy2008_prl,laurson10_pre}, and in experiments within a 2D interfacial configuration \cite{grob09_pag,maloy06_prl}: $\beta_{na}=1.7$. Moreover it is worth noting that this last exponent is clearly different from the one obtained from the acoustics emission in experiments. Acoustic emission are not directly related to the local depinning jumps of the fracture front. 

The inset of Fig. \ref{fig_richter}(c) shows that the threshold $v_{th}$, heuristically chosen to measure avalanches, does not change the value of $\beta$. Conversely, it significantly affects the upper cut-off $S_{max}$. This is shown here on the global $\overline{v}_{th}$ signal of the numerical simulation. This has been found to be true for the other measurement methods, on the different observables. This is even true for the other statistical laws presented in this paper: The signal thresholding used to define the avalanches only modify the power-laws cut-offs. Similarly it has been shown numerically on the global avalanches that $S_{min}$ increases with $c/k$ and $S_{max}$ decreases with $c/k$, leading to the disappearance of the power-law at high $c$ and low $k$ \cite{bares18_prb}.

\begin{figure}[!h]
\centering\includegraphics[width=0.8\columnwidth]{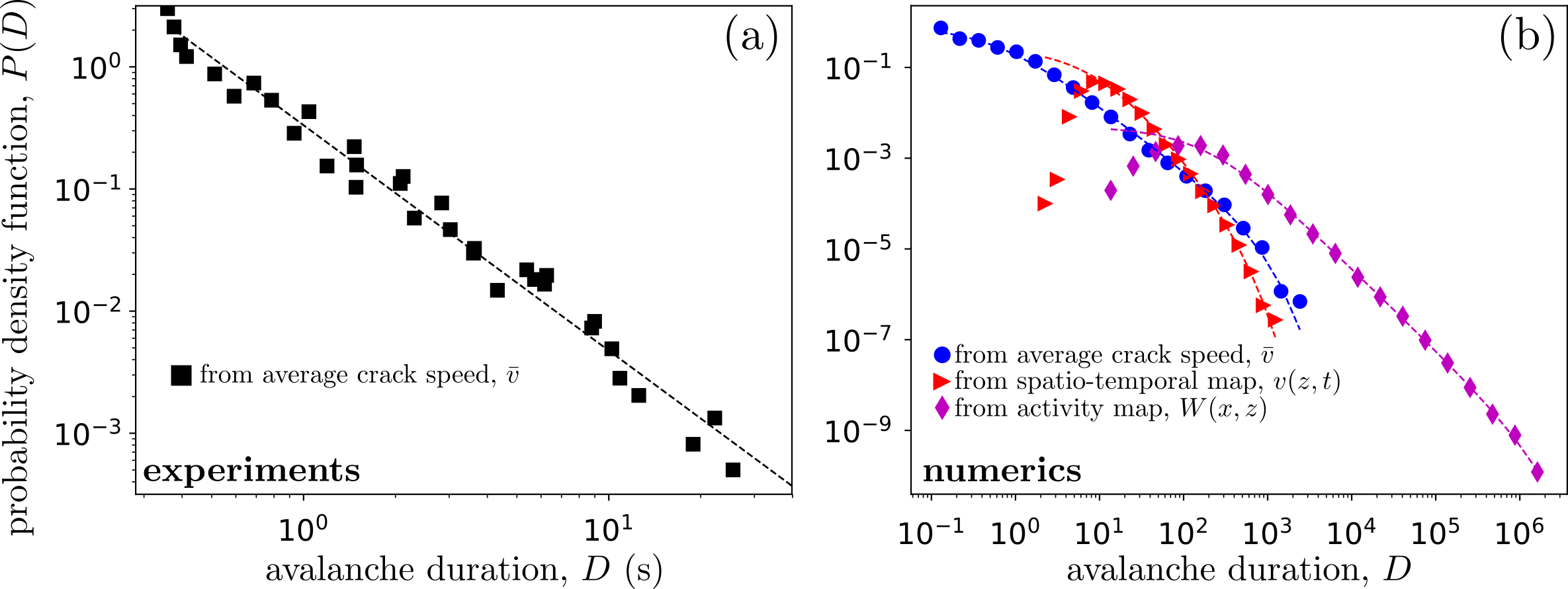}
\caption{Distribution of individual event duration $P(D)$ for experiments (panel a) and simulations (panel b). The different curves stand for different types of avalanches: events detected on the experimental $\overline{v}(t)$ signal (black $\Square$, panel a), events detected on the numerical $\overline{v}(t)$ signal (blue $\Circle$, panel b), events detected on the numerical spatio-temporal map $v(z,t)$ (red $\rhd$, panel b), events detected on the activity map $W(x,z)$ (purple $\Diamond$, panel b). All curves have been fitted using Eq. \ref{eq_PD} (dashed lines). The obtained fitting parameters are: $D_{min}^{eg}<<0.2$, $D_{max}^{eg}>>30$ and $\delta_{eg}=1.85\pm0.06$; $D_{min}^{ng}=0.56\pm0.16$, $D_{max}^{ng}=6.6\times10^{2}\pm0.9\times10^{2}$ and $\delta_{ng}=1.40\pm0.05$; $D_{min}^{nl}=9.7\pm4.6$, $D_{max}^{nl}=3.4\times10^{2}\pm0.8\times10^{2}$ and $\delta_{nl}=2.29\pm0.25$; $D_{min}^{na}=183\pm33$, $D_{max}^{na}=1.4\times10^{6}\pm0.3\times10^{3}$ and $\delta_{na}=1.80\pm0.03$. In panel a, the points are obtained by superimposing data from different avalanche detection threshold $\overline{v}_{th}$.}
\label{fig_duree}
\end{figure}

The avalanche duration $D$ also obeys power-law distribution, both in the experiment and simulation (Fig. \ref{fig_duree}). In the numerical case, the data are well fitted by the following PDF :

\begin{equation}
P(D) \sim \frac{e^{-D/D_{max}}}{(1+D/D_{min})^{\delta}},
\label{eq_PD}
\end{equation}

\noindent where $D_{min}$ and $D_{max}$ are the lower and upper cut-offs respectively and $\delta$ is the exponent of this gamma law. From the experimental side, $P(D)$ is a pure power-law without any cut-off when global avalanches are considered (Fig. \ref{fig_duree}(a)). The associated exponent is: $\delta_{eg}=1.85\pm0.06$. This value is significantly higher than the one measured in its numerical counterpart: $\delta_{ng}=1.40\pm0.05$. It has been shown, in \cite{bares18_prb}, that this exponent varies with $c$ (loading speed) and $k$ (unloading factor). Most likely, the $c$ and $k$ values prescribed in the numerical simulation do not correspond with the ones of the experiment so we do not expect $\delta_{eg}$ and $\delta_{ng}$ to be equal. Still $\delta_{ng}$ is close to the value expected for long-range depinning transition in the quasistatic limit, $\delta_g=1.5$ \cite{bonamy2009_jpd}. 

Regarding the local avalanches in the simulation (dug out from the $v(z,t)$ spatio-temporal map), the measured exponent is $\delta_{nl}=2.29\pm0.25$. A significantly lower value is obtained when the local avalanches are detected from the $W(x,z)$ activity map: $\delta_{na}=1.80\pm0.03$ (Fig.\ref{fig_duree}(b)). We also note that the avalanche duration measured acoustically on the experiment is meaningless since, due to wave reverberation, it depends on the sample geometry. 

\begin{figure}[!h]
\centering\includegraphics[width=0.8\columnwidth]{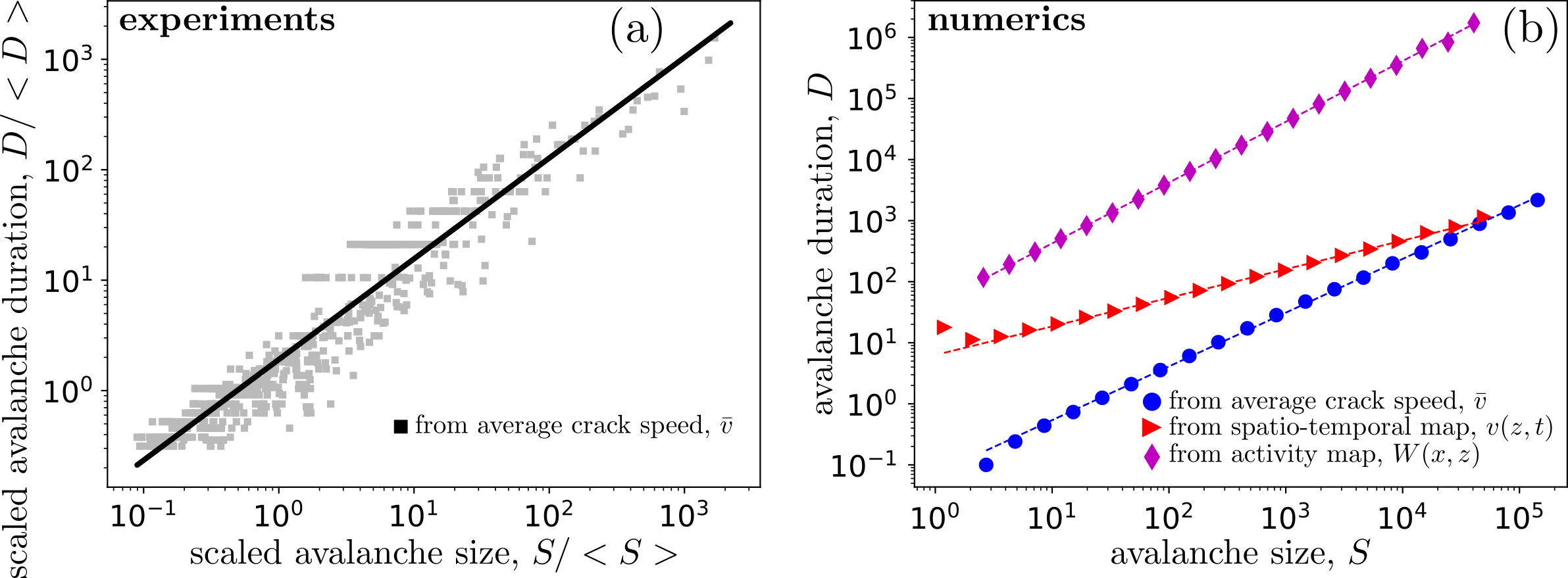}
\caption{Scaling between avalanche size $S$ and duration $D$ for experiment (panel a) and simulation (panel b). The different curves stand for the different types of avalanches: Global events detected on the experimental $\overline{v}(t)$ signal (black $\Square$, panel a), global events detected on the numerical $\overline{v}(t)$ signal (blue $\Circle$, panel b), local events detected on the numerical $v(z,t)$ spatio-temporal map (red $\rhd$, panel b), local events detected on the $W(x,z)$ activity map (purple $\Diamond$, panel b). All points have been fitted by power-laws (straight lines). The obtained exponents are: $\gamma_{eg}=0.91\pm0.01$, $\gamma_{ng}=0.880\pm0.006$, $\gamma_{nl}=0.470\pm0.003$ and $\gamma_{na}=0.996\pm0.003$.}
\label{fig_SvsD}
\end{figure}

Figure \ref{fig_SvsD} presents the scaling of avalanche size, $S$, with duration, $D$. Regardless of the type of avalanche considered, one gets:

\begin{equation}
D \sim S^{\gamma} 
\label{eq_SD}
\end{equation}

\noindent Experimentally and with regard to global avalanches, the exponent is $\gamma_{eg}=0.91\pm0.01$ (Fig. \ref{fig_SvsD}(a)). Avalanches were obtained using different detection thresholds $v_{th}$ and, as such, $S$ and $D$ have been rescaled by their respective mean values so that all curves collapse onto a single master one. This experimental exponent is found to be very close to the one observed in the simulation (Fig. \ref{fig_SvsD}(b)): $\gamma_{ng}=0.880\pm0.006$. These two exponents are however significantly higher than that at the critical point for a long-range depinning transition in the quasi-static limit (that is $c \rightarrow 0$, $k \rightarrow 0$): $\gamma=0.55$ \cite{bonamy2009_jpd}. They are also higher than the values $0.55-0.7$ reported in 2D interfacial crack experiments \cite{Laurson13_natcom,janicevic2016_prl}. For local avalanches detected from the $W(x,z)$ activity maps and on $v(z,t)$  spatio-temporal maps, the exponents are different: $\gamma_{na}=0.996\pm0.003$ in the case of activity maps and $\gamma_{nl}=0.470\pm0.003$ in the case of spatio-temporal maps, that is  about half the exponent measured for global avalanches.

\begin{figure}[!h]
\centering\includegraphics[width=0.85\columnwidth]{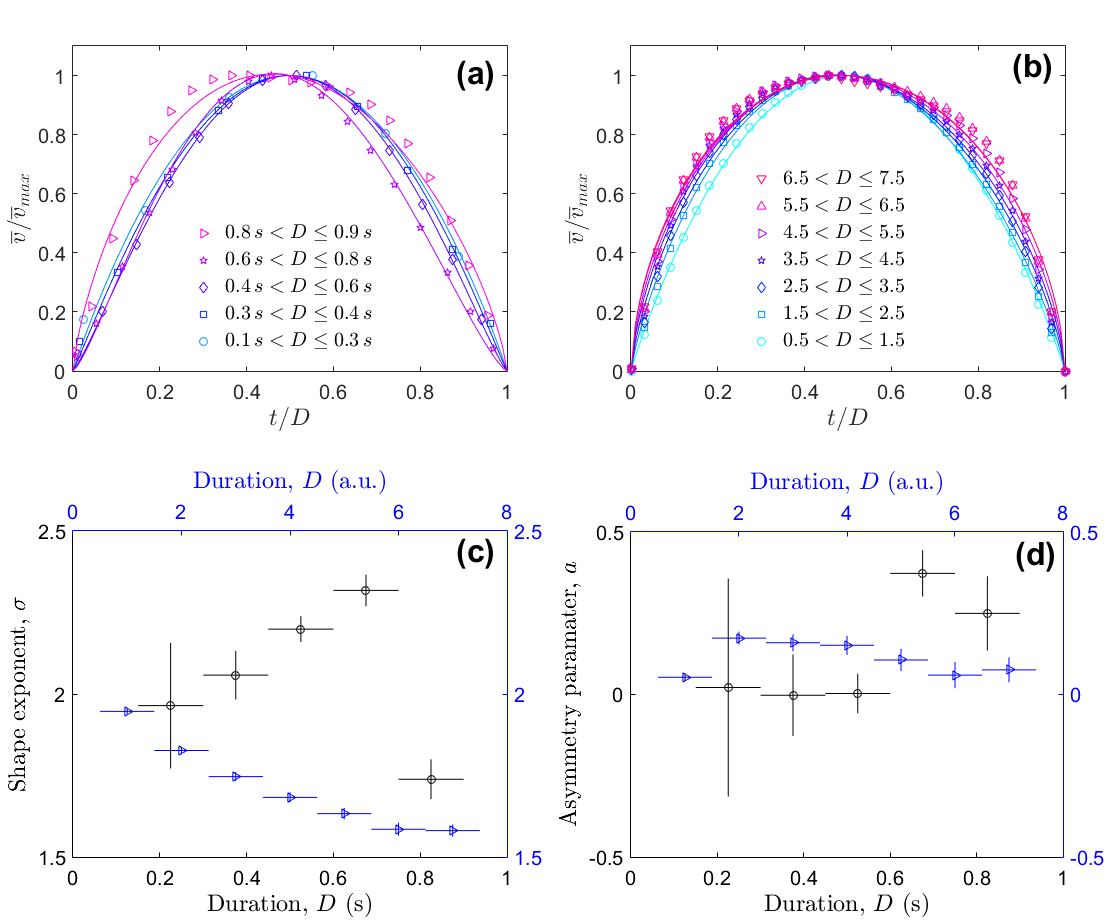}
\caption{Avalanche shapes extracted from averaged crack speed $\overline{v}(t)$ for experiments (panel a) and simulations (panel b). The duration $D$ of the avalanche collected to measure the shape are varied from $0.2$~s to $0.9$~s in the experimental case and $1$ to $7$ in the numerical case. In both panels (a) and (b), the markers are the measured shapes and the lines are fits using Eq. \ref{eq_Shp}. The fitted exponent $\sigma$ and asymmetry parameter $a$ are plotted as a function of $D$ in panels (c) and (e). In panels (c) and (d), blue symbols $\triangleright$ correspond to the simulation while black symbols $o$ correspond to experiment. Errorbars show a $95\%$ confident interval.}
\label{fig_forme}
\end{figure}

Finally, we have characterized the temporal shape of the global avalanches, and their evolution with $D$ (Fig. \ref{fig_forme}). This observable, indeed, provides an accurate characterization of the considered crackling signal and, as such, has been measured experimentally and numerically in a variety of systems \cite{zapperi05_nat,mehta06_pre,laurson06_pre,Papanikolao11_nat,Danku13_prl,Laurson13_natcom,bares14_prl}. The standard procedure was adopted here: First, we identified all avalanches with durations $D_i$ falling within a prescribed interval $\left[ D-\varepsilon , D+\varepsilon \right]$; and second, we averaged the shape $\overline{v}(t|D)/\max_{t \in \left[ t_{i}^{start} , t_{i}^{end} \right]} (\overline{v}(t|D)) , t \in \left[ t_{i}^{start} , t_{i}^{end} \right]$ over all the collected avalanches. Figures \ref{fig_forme}(a) and \ref{fig_forme}(b) show the resulting shape, for the experiment and simulation. We observe in both case that the shape is nearly parabolic at small $D$ with a very small asymmetry. The shapes were fitted using the scaling form proposed in \cite{Laurson13_natcom}:  

\begin{equation}
\left\langle\frac{\overline{v}(t|D)}{\max(\overline{v}(t|D)} \right\rangle=\left[4 \frac{t}{D}\left(1-\frac{t}{D}\right)\right]^{\sigma-1}\left[1-a\left(\frac{t}{D}-\frac{1}{2}\right)\right],
\label{eq_Shp}
\end{equation}

\noindent where $\sigma_{eg}$ (resp. $\sigma_{ng}$) is the shape exponent and $a_{eg}$ (resp. $a_{ng}$) quantifies the shape asymmetry in the experiment (resp. in the simulation). At small $D$, $\sigma_{eg} \approx \sigma_{ng} \approx 2$ which is consistent with a parabolic shape. We note that the prediction \cite{Laurson13_natcom} $\sigma=1/\gamma$ is not fulfilled in our case, neither in the experiment nor in the simulation. This may be due to the combined effects of a finite driving rate and a finite threshold value, yielding both overlaps between the depinning avalanches \cite{bares13_prl} and the splitting of depinning avalanches into separate sub-avalanches \cite{janicevic2016_prl}); neither of these effects are taken into account in the analysis proposed in \cite{Laurson13_natcom}. We also note that $\sigma$ evolves with $D$: It increases with increasing $D$ in the experiment and decreases with increasing $D$ in the simulation (Fig. \ref{fig_forme}(c)). We finally note that the visual flattening observed in Figs. \ref{fig_forme}(a) and \ref{fig_forme}(b) is captured less and less by the scaling form \ref{eq_Shp} as $D$ gets large. Similar features were observed in Barkhausen pulses \cite{Papanikolao11_nat} and was shown to result from the finite value of the demagnetization factor. The same is to be expected here since the unloading factor $k$ in Eq. \ref{eqLine} plays the same role as the demagnetization factor in the Barkhausen problem \cite{bares14_prl}. Finally a small but clear leftward asymmetry is detected (positive $a$ in Fig. \ref{fig_forme}(c)): The bursts start faster than they stop. We note that it is the opposite of what is observed for plasticity avalanches in amorphous materials \cite{liu16_prl} and consistent with that observed in \cite{Laurson13_natcom}. The asymmetry is much more pronounced in experiments than in the simulations. We conjectured \cite{bares14_prl} that it results from the viscoelastic nature of the polymer rock fractured here, which provides a negative inertia to the crack front, that is the addition of a retardation term in the dynamics equation \ref{eqLine} which was demonstrated \cite{zapperi05_nat}, in the Barkhausen context, to yield a significant leftward asymmetry in the pulse shape.

\section{Time-size organization of the event sequences}

\begin{figure}[!h]
\centering\includegraphics[width=\columnwidth]{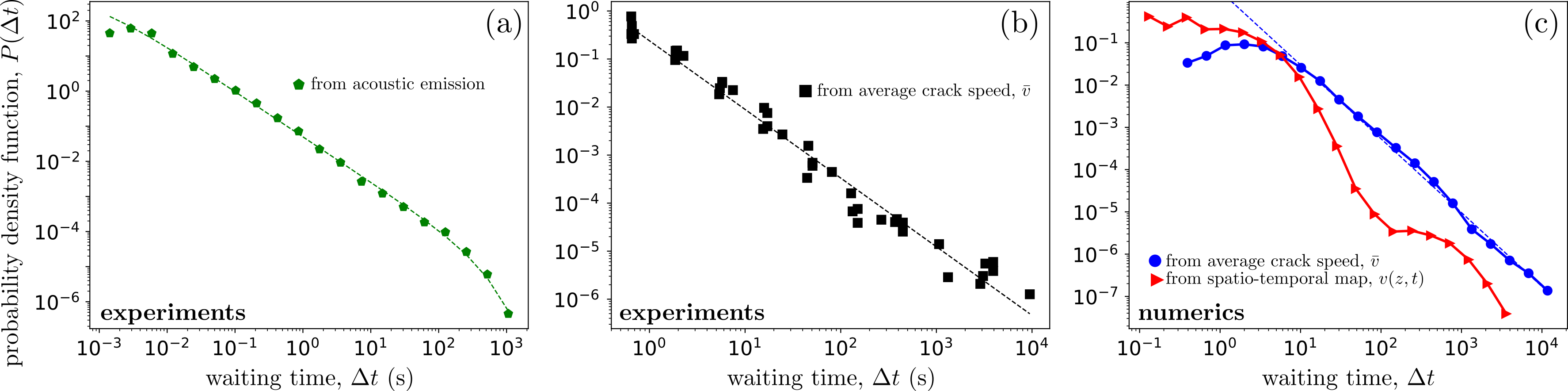}
\caption{Distribution of waiting time, $\Delta t$, between two consecutive events for experiments (panel a and b) and simulations (panel c). The different curves stand for different types of avalanches: acoustic events (green $\pentagon$, panel a), events detected on the experimental $\overline{v}(t)$ signal (black $\Square$, panel b), events detected on the numerical $\overline{v}(t)$ signal (blue $\Circle$, panel c), events detected on the numerical spatio-temporal map $v(z,t)$ (red $\rhd$, panel c). In panel a, the dashed line is a gamma-law fit with exponent $p_{ea}=1.29\pm0.02$ and upper cut-off $\Delta t_{max}=4.35\times10^{2}\pm0.49\times10^{2}$. In panel b and c, the curves corresponding to the avalanches detected on the $\overline{v}(t)$ signal have been fitted by a power-law (straight dashed lines). The fitted exponents are: $p_{eg}=1.43\pm0.03$ and $p_{ng}=1.75\pm0.03$.}
\label{fig_attente}
\end{figure}

We now turn to the statistical organization of the successive events, beyond their individual scale-free statistics. Regarding global avalanches, the recurrence time, $\Delta t$, is power-law distributed in both the experiments (Fig. \ref{fig_attente}(a)) and simulations (Fig. \ref{fig_attente}(b)). In both cases, the associated exponents, $p_{eg}$ (experiments) and $p_{ng}$ (numerics) are not universal; they significantly evolve with the mean crack speed \cite{bares13_phd,bares18_prb}. Since there is no one-to-one relation between the experimental and numerical control parameters, we cannot comment further on the difference between $p_{eg}$ and $p_{ng}$.

Experimentally, the waiting time separating two successive acoustic events is also power-law distributed (Fig. \ref{fig_attente}(b)). The associated exponent, $p_{ea}$, is significantly smaller than $p_{eg}$: $p_{ea} \simeq 1.16$ for $\overline{v}=2.7~\mu\mathrm{m/s}$, to be compared to $p_{eg} \simeq 1.76$ in the same experiment. Note also that, $p_{ea}$, as $p_{eg}$, significantly depends on $\overline{v}$ \cite{bares2018_natcom}. Experiments performed in artificial rocks made from beads of smaller sizes ($d=24,\mu\mathrm{m}$ or $d=233,\mu\mathrm{m}$) have also revealed that $p_{ea}$ depends on the microstructural length-scale \cite{bares2018_natcom}. Back to numerical simulations, the analysis of the local avalanches identified from the statio-temporal maps does not reveal any special time correlation; the waiting time is not scale-free (Fig. \ref{fig_attente}(c)). This suggests that the time correlation evidenced in the global avalanches emerges from the time overlapping of the local avalanches. Note that the time clustering evidenced here in the acoustic emission (as well as its absence with respect to local avalanches in the simulation) is visually reflected in the spatio-temporal map shown in Fig. \ref{fig_display}(g) (resp. in that shown in Fig. \ref{fig_display}(b)), with acoustic events gathered in time bands (resp. numerical avalanches  distributed randomly).

\begin{figure}[!h]
\centering\includegraphics[width=0.45\columnwidth]{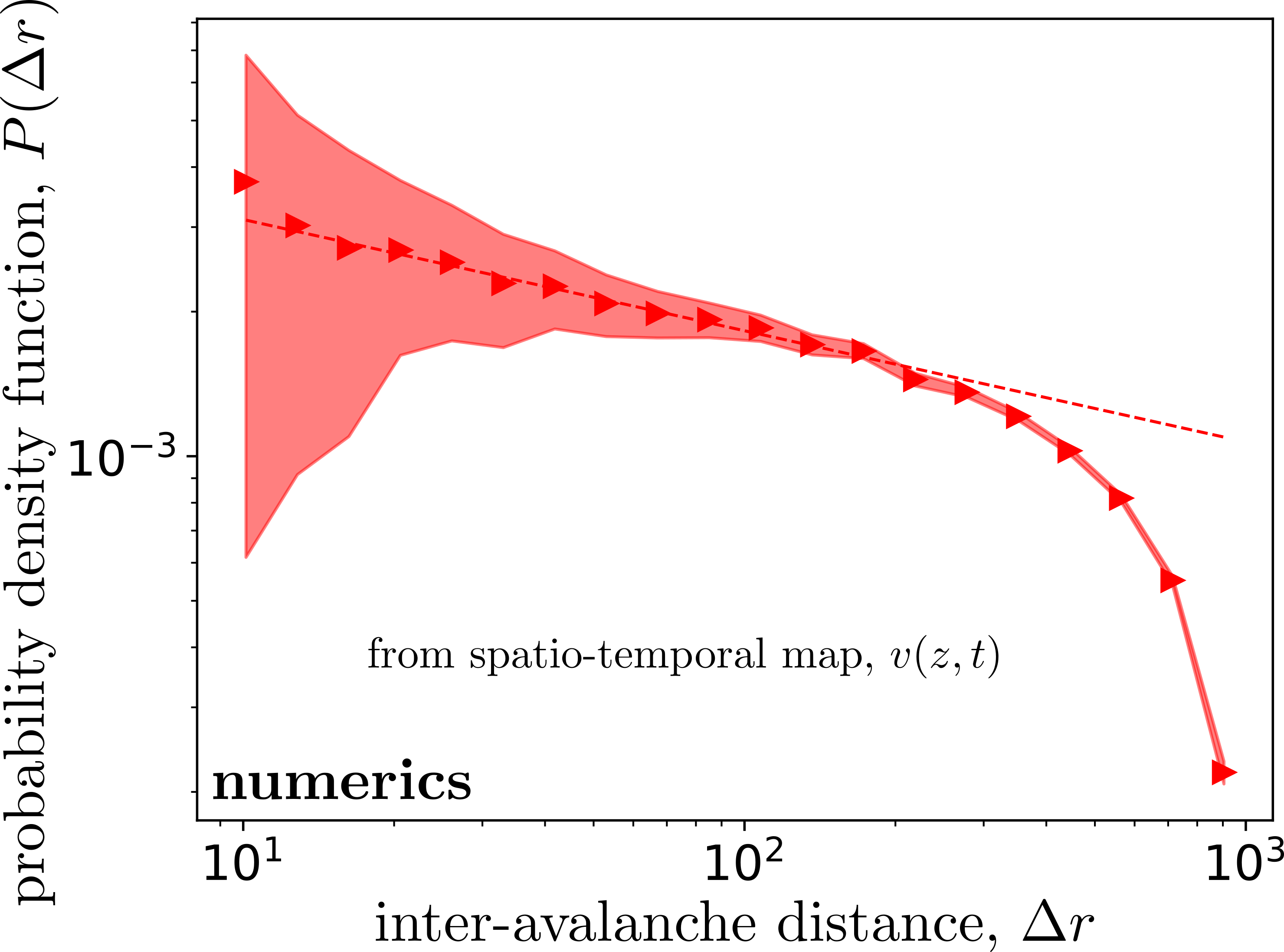}
\caption{Distribution of distances, $\Delta r$, between two consecutive local events detected on the spatio-temporal map $v(z,t)$. The red shaded area shows the $95$\% errorbar. The straight dashed line is a power-law fit with exponent $\lambda = 0.23\pm0.01$.}
\label{fig_space_clustering}
\end{figure}

In this context, it is of interest to look at the distribution of inter-event distances, $\Delta r$, for the local avalanches identified in the space-time maps (Fig. \ref{fig_space_clustering}). These statistics are found to be power-law distributed:

\begin{equation}
P(\Delta r) \sim \Delta r^{-\lambda},
\label{eq_davidsen}
\end{equation}

\noindent with an associated exponent $\lambda \simeq 0.23$. Similar scale-free statistics are observed in seismicity catalog \cite{davidsen2013_prl}, or in lab scale experiments driving a tensile crack front along an heterogeneous interface \cite{grob09_pag}. In both these cases, the value $\lambda$ is reported to be significantly larger than that measured here, around $0.6$.

The time correlation evidenced above, for global and acoustic events, are reminiscent of what is observed in earthquakes \cite{bak02_prl}, or during the gradual damaging of heterogeneous solids under compressive loading conditions \cite{baro13_prl,makinen2015_prl}. In both these situations, the events are known to form characteristic aftershock (AS) sequences obeying specific scaling laws: Productivity law \cite{helmstetter2003_prl,utsu1971_jfs} telling that the number of produced AS goes as a power-law with the mainshock (MS) size; B\r{a}th's law stating that the ratio between the MS size and that of its largest AS is independent of the MS magnitude, and Omori-Utsu law stipulating that the production rate of AS decays algebraically with time to MS. Hence, for each type of events, we have decomposed the series into aftershock sequences and analyzed them at the light of these laws. 

In the seismology context, many different clustering methods \cite{stiphout2012_com} have been set-up to separate the AS sequences. Most of them are based on the proximity between events, in both time and space. Unfortunately, spatial proximity is not relevant here, because of the lack of information on the event position for global and acoustic events (Tab. \ref{tab_method}). Hence, we have chosen the method developed in \cite{baro13_prl,bares2018_natcom}, which makes use of the  occurrence time $t_i$ only. The procedure is the following: First, a size $S_{MS}$ is prescribed and all events of size falling within the interval $S_{MS} \pm \delta S_{MS}$ are labeled as MS; second, for each MS, all subsequent events are considered as AS, until an event of size larger than that of the MS is encountered. From the numerical side, the analysis has been performed on both the global avalanches (dug up from $\overline{v}(t)$) and the local ones (dug up from the space-time map $v(z,t)$). From the experimental side, the analysis has been performed on the acoustic events. conversely, It could not have been achieved on the global experimental events, due to a lack of statistics (few hundreds of events only). 

\begin{figure}[!h]
\centering\includegraphics[width=0.8\columnwidth]{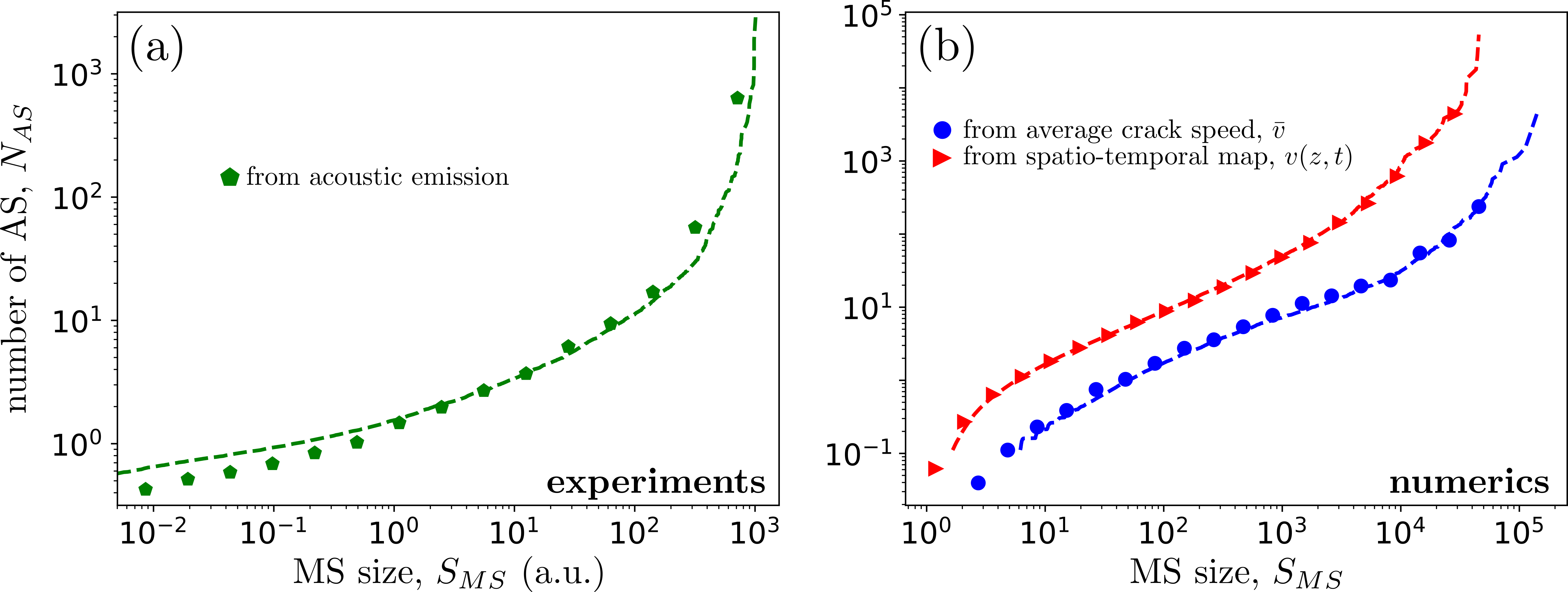}
\caption{Mean AS number, $N_{AS}$, as a function of the triggering MS size, $S_{MS}$, for experiments (panel a) and simulations (panel b). The different curves stand for the different types of avalanches: Acoustic events (green $\pentagon$, panel a), global events detected on the numerical $\overline{v}(t)$ signal (blue $\Circle$, panel b), local events detected on the numerical spatio-temporal map $v(z,t)$ (red $\rhd$, panel b). In each case, the dashed line is given by Eq. \ref{eq_prod}.}
\label{fig_productivite}
\end{figure}

Figure \ref{fig_productivite} shows the mean number of AS, $N_{AS}$, triggered by a MS of size $S_{MS}$, for acoustic events (panel a) and global/local avalanches in the simulation (panel b). In the three cases, the productivity law is fulfilled and there is a range of decades over which $N_{AS}$ scales as a power-law with $S_{MS}$. Actually, such a behavior has been demonstrated \cite{bares2018_natcom} to emerge naturally from the scale-free statistics of size; calling $F(S)=\int_{S_{min}}^S P(S) \ud S$ the cumulative distribution of size, the total number of events in the series to be labeled AS is $F(S_{MS})$ and the total number of MS -- hence AS sequences -- is $1-F(S_{MS})$. Hence, the mean number per AS sequence is the ratio between the two:  

\begin{equation}
N_{AS}(S_{MS})=\frac{F(S_{MS})}{1-F(S_{MS})},
\label{eq_prod}
\end{equation}

\noindent which fits perfectly the data, without any adjustable parameter. Note that,for a pure scale-free statistics $P(S) \sim S^{-\beta}$, Eq. \ref{eq_prod} would have yielded $N_{AS} \sim S_{MS}^{\beta-1}$.  In other words, it is the presence of finite lower and upper cutoffs, $S_{min}$ and $S_{max}$, which is responsible for the departure to this pure power-law scaling.    

\begin{figure}[!h]
\centering\includegraphics[width=0.8\columnwidth]{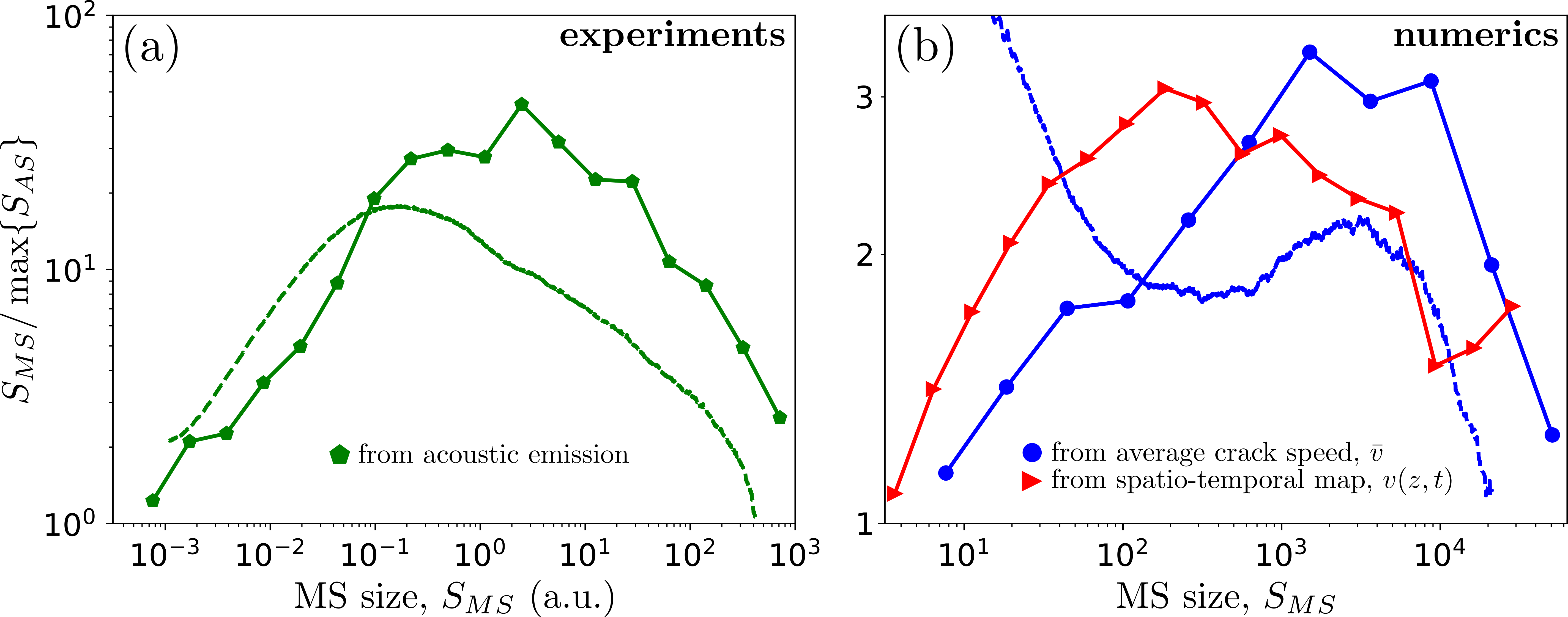}
\caption{Mean size ratio, $S_{MS}/\max\{S_{AS}\}$, between a MS and its largest AS for experiments (panel a) and simulations (panel b). The different curves stand for the different types of avalanches: acoustic events (green $\pentagon$, panel a), global events detected on the numerical $\overline{v}(t)$ signal (blue $\Circle$, panel b), local events detected on the numerical spatio-temporal map $v(z,t)$ (red $\rhd$, panel b). In each case, the dashed line is given by Eq. \ref{eq_bath}.}
\label{fig_bath}
\end{figure}

B\r{a}th's law relates the largest AS size in the sequence to that of the triggering MS; it states that the ratio between the two is independent of the MS size. This ratio $S_{MS}/\max\{S_{AS}\}$ is plotted as a function of $S_{MS}$ in Fig. \ref{fig_bath} for the experiments (acoustic events) and simulations (global and local avalanches). As for the productivity law, a simple prediction can be obtained by considering independent events whose distribution in size is $P(S)$. One can then use extreme event theory to derive the statistical distribution of a largest event of size $S$ in a sequence with $N_{AS}$ AS \cite{bares2018_natcom}. The mean value of this maximum value follows \cite{bares2018_natcom}: 

\begin{equation}
\frac{\max(S_{AS})}{S_{MS}}=N_{AS}(S_{MS})\times \int_{S_{min}}^{S_{MS}} S F(S)^{N_{AS}(S_{MS})-1}P(S)\ud S, 
\label{eq_bath}
\end{equation}

\noindent where $N_{AS}(S_{MS})$ is given by Eq. \ref{eq_prod}, P(S) is given by Eq. \ref{eq_PS}, and $F(S)=\int_{S_{min}}^S P(S)\ud S$.  

Finally, Omori-Utsu law was addressed. For each type of events, the number of AS per unit time, $r_{AS}(t|S_{MS})$, is computed by binning the AS events over $t-t_{MS}$ and subsequently averaging the so-obtained curves over all MS with size falling into the prescribed interval $(1\pm\epsilon)S_{MS}$. In all cases, the algebraic decay expected from the Omori-Utsu law is observed. The prefactor increases with $S_{MS}$, which is expected since $N_{AS}$ increases with $S_{MS}$ (Eq. \ref{eq_prod}). It has been reported in \cite{bares2018_natcom} that, for acoustic events, all curves can be collapsed by dividing time by $N_{AS}(S_{MS})$, so that the overall production rate writes:   

\begin{equation}
r_{AS}(t|S_{MS})=f\left(\frac{t-t_{MS}}{N_{AS}(S_{MS})}\right)\quad \mathrm{with} \quad f(u)\sim \frac{e^{-u/\tau_{max}}}{(1+u/\tau_{min})^p}
\label{eq_omori}
\end{equation}

\begin{figure}[!h]
\centering\includegraphics[width=0.8\columnwidth]{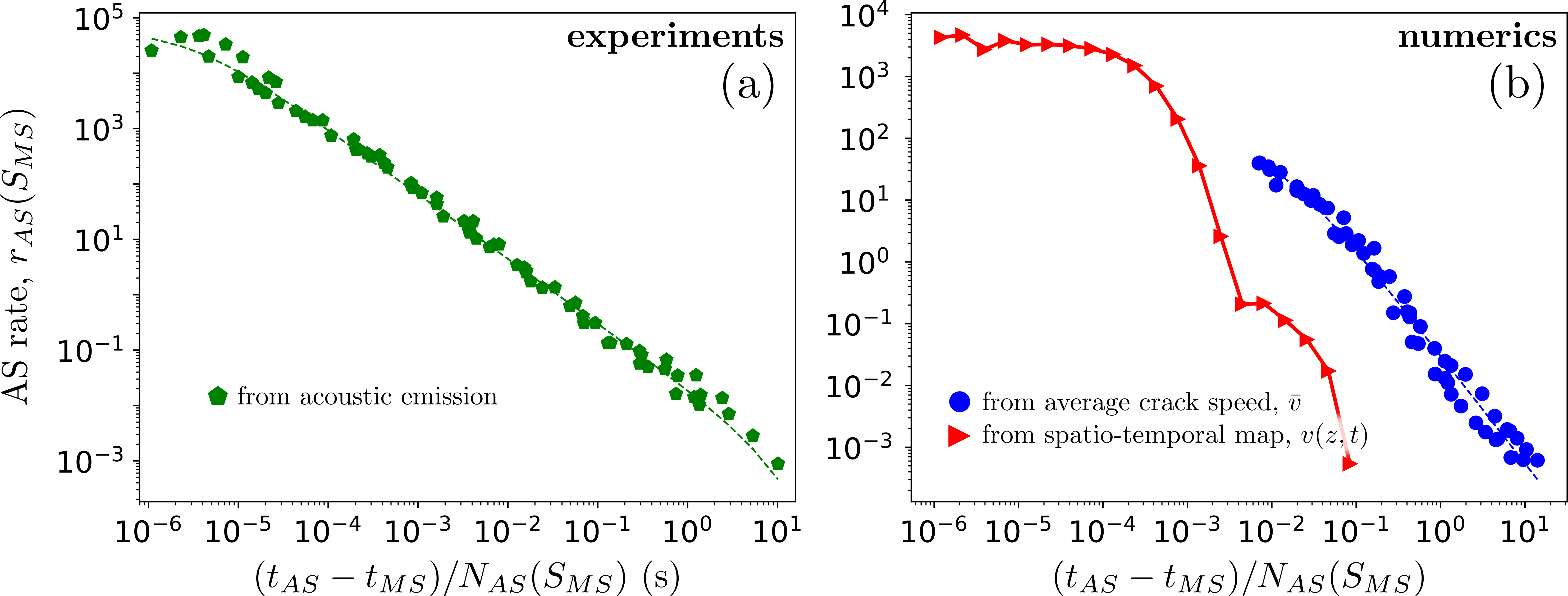}
\caption{AS rate, $r_{AS}$, as a function of the time elapsed since MS, $t_{AS}-t{MS}$, for experiments (panel a) and simulations (panel b). The curves are scaled by the productivity $N_{AS}$ as proposed by Eq. \ref{eq_omori}. The different curves stand for the different types of avalanches: acoustic events (green $\pentagon$, panel a), events detected on the numerical $\overline{v}(t)$ signal (blue $\Circle$, panel b), events detected on the numerical spatio-temporal map $v(z,t)$ (red $\rhd$, panel b). All curves have been fitted using Eq. \ref{eq_omori} (dashed lines). The obtained fitting parameters are: $\tau_{min}^{ea}=2.83\times10^{-6}\pm1.36\times10^{-6}$, $\tau_{max}^{ea}=9.3\pm5.2$ and $p_{ea}=1.17\pm0.02$; $\tau_{min}^{ng}=8.45\times10^{-3}\pm4.86\times10^{-3}$, $\tau_{max}^{ng}>>10$ and $p_{ng}=1.75\pm0.11$. In the experimental case (panel a) the points are obtained by superimposing data with different $S_{MS}$. In all cases, the avalanche size threshold is fixed $S_{th}=0$.}
\label{fig_omori}
\end{figure}

\noindent This collapse is verified here, not only for acoustic events (Fig. \ref{fig_omori}(a)), but also for the global avalanches in simulations (Fig. \ref{fig_omori}(b)). It has also been demonstrated on AE \cite{bares2018_natcom} that the Omori-Utsu exponent, $p$, is the same as that of $P(\Delta t)$. This is found to be true for the global avalanches, also. Let us finally mention that Eq. \ref{eq_omori} is not fulfilled for the local avalanches detected onto the space-time numerical maps (Fig. \ref{fig_omori}(b)); this is coherent with the fact that inter-event times were not scale-free for this type of avalanches, neither. 

\section{Concluding discussion}

We examined here the crackling dynamics in nominally brittle crack problem. Experimentally, a single crack was slowly pushed into an artificial rock made of sintered polymer beads. An irregular burst-like dynamics is evidenced at the global scale, made of successive depinning jumps spanning a variety of sizes. The area swept by each of these jumps, their duration, and the overall energy released during the event is power-law distributed, over several orders of magnitude. Despite their individual giant fluctuations, the ratio between instantaneous, spatially-averaged, crack speed and power release remains fairly constant and defines a continuum-level scale material constant fracture energy. 

The features depicted above can be understood in a model which explicitly takes into account the microstructure disorder by introducing a stochastic term into the continuum fracture theory. Then, the problem of crack propagation maps to that of a long-range elastic interface driven by a force self-adjusting around the depinning threshold. This approach reproduces the crackling dynamics observed at global scale. The agreement is quantitative regarding size distribution; the exponents measured experimentally and numerically are very close. They are also very close to the value $\beta_{g}=1.28$ predicted theoretically via Functional Renormalization Group (FRG) method \cite{bonamy2009_jpd,ledoussal09_pre}. Conversely, the exponent characterizing the scale-free statistics of the event duration, $\delta_g$, is different in the experiment and in the simulation. The former is rather close to the predicted FRG value, $\delta_g=1.50$. Note that FRG analysis presupposes a quasi-static process, with a vanishing driving rate (parameter $c$ in Eq. \ref{eqLine} and simulation, $V_{wedge}$ in the experiment). By yielding some overlap between the global avalanches, a finite driving rate may change the value of $\delta$ \cite{white03_prl,bares13_prl}. Different driving rates in the experiment and simulation may also be at the origin of the difference between $\delta_{eg}$ and $\delta_{ng}$. Note also that the long-range elastic kernel in Eq. \ref{eq_mod0} is actually derived assuming infinite thickness. This may not be relevant in our experiment where the specimen thickness is only 30 times larger than the microstructure length-scale. In this respect, it is worth to note that values $\sim 1.5$ were experimentally measured in interfacial growth experiments with ratios thickness over microstructure scale much larger \cite{janicevic2016_prl}, i.e. more in line with the long range elastic kernel of Eq. \ref{eq_mod0}. 

The analysis of the simulations has permitted to define avalanches at the local scale, as localized depinning events in both space and time (in contrast with the global avalanches identified with $\overline{v}(t)$ bursts localized in time only). Two definitions were proposed: digging out these local avalanches either from activity map $W(x,z)$ or from space-time velocity map $v(z,t)$. Both cases lead to similar, scale-free, statistics for avalanche size; the two procedure are conjectured to be equivalent. Conversely, the obtained  exponent, $\beta_l \simeq 1.65$, are significantly higher than those associated with global avalanches. This illustrates that local and global avalanches are distinct entities; each global avalanche is actually made of numerous local avalanches \cite{laurson10_pre}. Unfortunately, the statistics of these local avalanche could not be determined in our experiments. Conversely, the value observed here is very close to that reported in interfacial crack experiments \cite{maloy06_prl,grob09_pag}. 

This global crackling dynamics goes along, in the experiment, with the emission of numerous acoustic events which are also power-law distributed in energy. The associated exponent, $\beta_{ea}\simeq 1$, is significantly smaller than those associated with global or local avalanche size. Actually, AE are elastodynamics quantities different from the depinning (elastostatic) avalanches: They are the signature of the elastic waves triggered by the local accelerations/decelerations within the depinning events, but their energy is not proportional to the depinning area (or to the total elastostatic energy released during the depinning). In particular, the acoustic waveform will depend not only on the depinning event, but also on the complete geometry of the specimen at the time of the event, the eigenmodes at that time, etc. Quite surprisingly, the size of the global avalanches (that is the length of the crack jump caused by a depinning event) has been observed \cite{bares2018_natcom} to be proportional to the number of acoustic events produced during the event rather than to the sum of acoustic energy cumulated over the event as was initially proposed in \cite{Stojanova14_prl}. Deriving the rationalization tools to infer the relevant information on the underlying depinning event from the analysis of the acoustic waveform provide a tremendous challenge for future investigation. 

Beyond their individual scale-free features, the acoustic events get organized in time and form characteristic AS sequences obeying the fundamental laws of seismicity: The productivity law relating the number of produced AS with the triggering MS size; B\r{a}th's law relating the size of the largest AS to that of the triggering MS and the Omori-Utsu law relating the AS production rate to the time elapsed since MS. These laws were recently demonstrated \cite{bares2018_natcom} to be a direct consequence of the individual scale-free statistics for size (for the productivity and B\r{a}th's law) and the scale-free statistics of inter-event time (for Omori-Utsu law). The sequences of global avalanches also obey similar time and size organization. In this context, the observation of Omori-Utsu law and scale-free statistics of inter-event times may appear surprising. Depinning models usually predict that, at vanishing driving rate, depinning events are randomly distributed, with an exponential distribution for inter-event time \cite{sanchez2002_prl}. However, it has been recently shown \cite{janicevic2016_prl} how the application of a finite threshold to identify the pulses in $\overline{v}(t)$ splits each true depinning avalanches into disconnected sub-avalanches with power-law distributed inter-event time. Note that, in this scenario, the characteristic exponent of the inter-event time is equal to that of the individual event duration, which is not observed here (Tab. \ref{tab_exponent}). This may result from a difference in the definition of the inter-event time, given by the difference in starting time between two successive events in our case, and by the difference between the starting time of an event and the ending time of its predecessor in \cite{janicevic2016_prl}. It is also interesting to note that local avalanches, in the simulation, do not display scale-free statistics for the inter-event times. Work in progress aims at understanding how such a scale-free statistics emerge at the global scale from the coalescence of the local avalanches at finite driving rate \cite{bares18_prb}.  

\begin{table}[]
\centering
\begin{tabular}{|c|c|c|c|c|}
\hline
statistics                                                                                       & observable                          & exponent       & value             & variability                    \\ \hline
\multirow{5}{*}{\begin{tabular}[c]{@{}c@{}}Richter-Gutenberg\\ $P(S)$\end{tabular}}              & from simulated $\overline{v}(t)$    & $\beta_{ng}$   & $1.36 \pm 0.05$   & \begin{tabular}{c} $\sim$ const. \cite{bares2014_ftp} \\ sligthly $\nearrow$ with $c$ \cite{bares18_prb} \end{tabular}                        \\ \cline{2-5} 
                                                                                                 & from simulated $v(z,t)$             & $\beta_{nl}$   & $1.62 \pm 0.03$   & $\sim$ const. \cite{bares13_phd}                               \\ \cline{2-5} 
                                                                                                 & from simulated activity $W(x,z)$    & $\beta_{na}$   & $1.66 \pm 0.05$   & $\sim$ const. \cite{bares13_phd,bonamy2008_prl}                               \\ \cline{2-5} 
                                                                                                 & from experimental $\overline{v}(t)$ & $\beta_{eg}$   & $1.35 \pm 0.10$   & slightly $\searrow$ with $\overline{v}$ \cite{bares13_prl} \\ \cline{2-5} 
                                                                                                 & from experimental acoustic          & $\beta_{ea}$   & $0.96 \pm 0.03$   & slightly $\searrow$ with $\overline{v}$ \cite{bares2018_natcom}\\ \hline
\multirow{4}{*}{\begin{tabular}[c]{@{}c@{}}Duration\\ $P(D)$\end{tabular}}                       & from simulated $\overline{v}(t)$    & $\delta_{ng}$  & $1.40 \pm 0.05$   & $\sim$ const. \cite{bares13_phd,bares2014_ftp}                               \\ \cline{2-5} 
                                                                                                 & from simulated $v(z,t)$             & $\delta_{nl}$  & $2.29 \pm 0.25$   &                                \\ \cline{2-5} 
                                                                                                 & from simulated activity $W(x,z)$    & $\delta_{na}$  & $1.80 \pm 0.03$   &                                \\ \cline{2-5} 
                                                                                                 & from experimental $\overline{v}(t)$ & $\delta_{eg}$  & $1.85 \pm 0.06$   &                                \\ \hline
\multirow{3}{*}{\begin{tabular}[c]{@{}c@{}}Waiting time\\ $P(\Delta t)$\end{tabular}}            & from simulated $\overline{v}(t)$    & $p_{ng}$       & $1.75 \pm 0.03$   & $\nearrow$ with $\overline{v}$ \cite{bares18_prb} \\ \cline{2-5} 
                                                                                                 & from experimental $\overline{v}(t)$ & $p_{eg}$       & $1.43 \pm 0.03$   &                                \\ \cline{2-5} 
                                                                                                 & from experimental acoustic          & $p_{ea}$       & $1.29 \pm 0.02$   & $\nearrow$ with $\overline{v}$ \cite{bares2018_natcom}\\ \hline
\begin{tabular}[c]{@{}c@{}}Jump\\ $P(\Delta r)$\end{tabular}                                     & from simulated $v(z,t)$             & $\lambda_{nl}$ & $0.23 \pm 0.01$   &                                \\ \hline
\multirow{2}{*}{\begin{tabular}[c]{@{}c@{}}Omori\\ $r_{AS}(t_{AS}-t_{MS})$\end{tabular}}         & from simulated $\overline{v}(t)$    & $p_{ng}$       & $1.75 \pm 0.11$   &                                \\ \cline{2-5} 
                                                                                                 & from experimental acoustic          & $p_{ea}$       & $1.17 \pm 0.02$   &                                \\ \hline
\multirow{4}{*}{S vs. D}                                                                         & from simulated $\overline{v}(t)$    & $\gamma_{ng}$  & $0.880 \pm 0.006$ & slightly $\searrow$ with $\overline{v}$ \cite{bares13_prl}                               \\ \cline{2-5} 
                                                                                                 & from simulated $v(z,t)$             & $\gamma_{nl}$  & $0.470 \pm 0.003$ &                                \\ \cline{2-5} 
                                                                                                 & from simulated activity $W(x,z)$    & $\gamma_{na}$  & $0.992 \pm 0.003$ &                                \\ \cline{2-5} 
                                                                                                 & from experimental $\overline{v}(t)$ & $\gamma_{eg}$  & $0.91 \pm 0.01$   &                                \\ \hline
\end{tabular}
\caption{Table of the exponents measured for the different statistical laws for avalanches detected on different numerical and experimental observables. In the subscript of the exponent names, `ng' stands for \textit{numerics global}, `nl' stands for \textit{numerics local}, `na' stands for \textit{numerics activity}, `eg' stands for \textit{experiment global} ans `ea' stands for \textit{experiment acoustic}.}
\label{tab_exponent}
\end{table}

\noindent \textbf{Acknowledgements}:

Support through the ANR project MEPHYSTAR (ANR-09-SYSC-006-01) and by "Investissements d'Avenir" LabEx PALM (ANR-10-LABX-0039-PALM). We thank Thierry Bernard for technical support, and Luc Barbier, Davy Dalmas and Alberto Rosso for fruitful discussions.


\end{document}